\documentclass[final,1p,times,twocolumn,authoryear]{elsarticle}

\usepackage{amssymb}
\usepackage{subfigure}
\usepackage{setspace}
\usepackage{amsmath}
\usepackage{amsfonts}
\usepackage{float}
\usepackage{multirow}
\usepackage{ulem} 
\usepackage{hyperref}

\usepackage{lineno}

\journal{ISPRS}

\begin{document}

\begin{frontmatter}



\title{ASANet: Asymmetric Semantic Aligning Network for RGB and SAR image land cover classification}


\author[inst1]{Pan Zhang\corref{cor1}\fnref{label1}}
\ead{whu\_zp@163.com}
\author[inst1]{Baochai Peng\fnref{label1}}
\ead{pengbaoc@gmail.com}
\author[inst1]{Chaoran Lu}
\author[inst1]{Quanjin Huang}
\author[inst1]{Dongsheng Liu}

\affiliation[inst1]{organization={PIESAT Information Technology Co Ltd.},
            city={Beijing},
            postcode={100195}, 
            country={China}}

\fntext[label1]{Equal contributions.}
\cortext[cor1]{Corresponding author}


\begin{abstract}
Synthetic Aperture Radar (SAR) images have proven to be a valuable cue for multimodal Land Cover Classification (LCC) when combined with RGB images. Most existing studies on cross-modal fusion assume that consistent feature information is necessary between the two modalities, and as a result, they construct networks without adequately addressing the unique characteristics of each modality. In this paper, we propose a novel architecture, named the Asymmetric Semantic Aligning Network (ASANet), which introduces asymmetry at the feature level to address the issue that multi-modal architectures frequently fail to fully utilize complementary features. The core of this network is the Semantic Focusing Module (SFM), which explicitly calculates differential weights for each modality to account for the modality-specific features. Furthermore, ASANet incorporates a Cascade Fusion Module (CFM), which delves deeper into channel and spatial representations to efficiently select features from the two modalities for fusion. Through the collaborative effort of these two modules, the proposed ASANet effectively learns feature correlations between the two modalities and eliminates noise caused by feature differences. Comprehensive experiments demonstrate that ASANet achieves excellent performance on three multimodal datasets. Additionally, we have established a new RGB-SAR multimodal dataset, on which our ASANet outperforms other mainstream methods with improvements ranging from 1.21\% to 17.69\%. The ASANet runs at 48.7 frames per second (FPS) when the input image is 256x256 pixels. The source code are available at \href{https://github.com/whu-pzhang/ASANet}{Github}.
\end{abstract}




\end{frontmatter}


\section{Introduction} \label{sec:Introduction}


LCC in remote sensing images is crucial for monitoring geological disasters \citep{li2014monitoring}, conducting land use analysis \citep{tu2020regional}, and facilitating urban planning \citep{guo2018semantic}. The advent of deep learning has significantly enhanced the intelligent interpretation of remote sensing images \citep{cao2023large}. Numerous existing methods (SLMFNet \citep{li2024slmfnet}, FURSformer \citep{zhang2023fursformer}, MF-DTCFCN \citep{li2022land}) primarily rely on single-source satellite data, such as optical imagery, for LCC. However, optical imagery is susceptible to interference from factors like clouds or adverse weather conditions during image acquisition, which can hinder its effectiveness in meeting practical application needs \citep{araya2018monitoring}.

Recent advancements in remote sensing technology have enabled the acquisition of data from various imaging modalities across the same geographic area. RGB images provide high spatial resolution and are rich in spectral and textural information; however, they are susceptible to weather conditions \citep{kazakeviciute2020assessment}. In contrast, SAR sensors function in all weather conditions and can penetrate certain types of ground cover \citep{ma2017review}, yielding detailed geometric information about ground objects. In specific applications, RGB and SAR images are highly complementary. The fusion of RGB and SAR images for remote sensing classification is considered a promising strategy to improve classification accuracy \citep{li2022deep}. Nonetheless, segmenting RGB and SAR images for remote sensing purposes presents greater challenges than multimodal segmentation in natural images \citep{zhang2020novel}.

\begin{figure*}[!h]
    \centering
    \includegraphics[width=\linewidth]{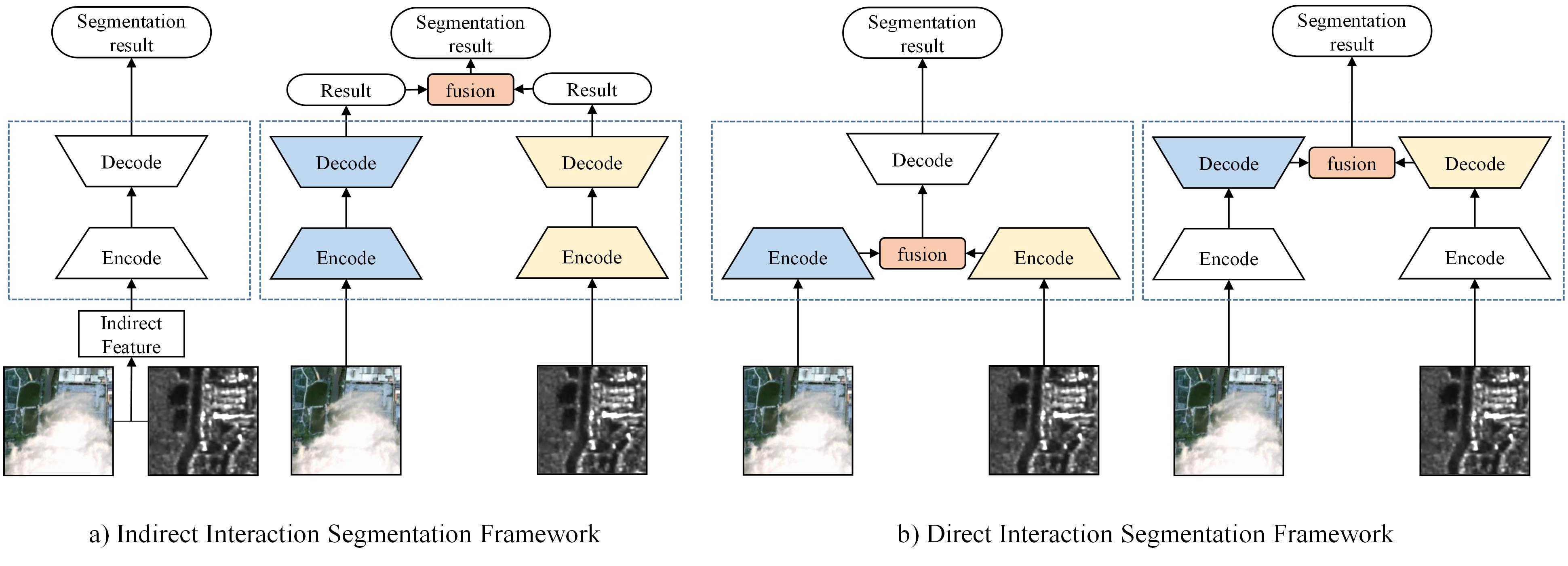}
    \caption{Comparison of different fusion frameworks: Features are categorized based on whether they engage in direct interaction within the network, being classified into two types: a) Indirect Interaction Segmentation Framework and b) Direct Interaction Segmentation Framework.}
    \label{fig:multimodel}
\end{figure*}

Previously, researchers frequently used machine learning techniques \citep{pei2014efficient,stone2013bayes} to post-process the segmentation results obtained from both imaging modalities. However, these post-processed results were generally not reusable for further network learning. Currently, deep learning-based multimodal semantic segmentation approaches (CMFNet \citep{ma2022crossmodal}, FTransUNet \citep{ma2024multilevel}, SwinTFNet \citep{ren2024swintfnet}) are increasingly utilized for LCC. As depicted in Fig. \ref{fig:multimodel}, these methods can be classified into two distinct types based on whether feature interaction occurs within the network architecture.

a) Feature Indirect Interaction Framework: This framework refers to a segmentation network that lacks direct feature interaction between two modalities during feature extraction or feature aggregation. The Align-CR \citep{xu2023high} is employed to generate an indirect image training classification model. \citep{hang2020classification} proposed a network that involves fusing the segmentation results from the two branches and then reintroducing them into the network for further learning. While these interaction methods contribute to reducing the discrepancies between modalities, their effectiveness is constrained by the lack of direct connections between modal features. This limitation impedes the network’s capability to learn complex feature associations.

b) Feature Direct Interaction Framework: This framework pertains to a segmentation network that promotes feature interaction between two modalities during feature extraction or feature aggregation. Initial methods, such as those proposed by \citep{ma2022amm}, \citep{kang2022cfnet}, and \citep{hazirbas2017fusenet}, generally resorted to simple feature concatenation or summation to integrate feature maps from the two modalities. However, these basic fusion techniques could inadvertently incorporate irrelevant information, potentially disrupting the network’s learning process. To address this issue, techniques like SA-Gate \citep{chen2020bi} and MCANet \citep{li2022mcanet} have introduced attention mechanisms to compute feature weights between the modalities, thus enabling more effective feature selection and fusion. These symmetric fusion frameworks operate under the assumption that key features are consistently shared across both modalities. Nonetheless, the distinct imaging mechanisms of RGB and SAR suggest inherent differences between the modalities. The traditional symmetric fusion method assigns the same attention weight to different branches, making the features of the two modalities appear closer. However, this approach hinders the learning of complementary features between different modalities.

To fully leverage the complementary features and establish robust feature correlations between RGB and SAR images, we design an Asymmetric Semantic Aligning Network (ASANet). This network is composed of two key modules: the Semantic Focusing Module (SFM) and the Cascade Fusion Module (CFM). The SFM captures the unique characteristics of RGB and SAR images by assigning different attention weights to features during their interaction process. The CFM enables the network to grasp the complex relationships between the two modalities by learning from both channel and spatial perspectives, thereby facilitating the fusion of features from the two modalities. By integrating the SFM and CFM, our network can effectively extract and correlate features from both modalities, significantly diminishing the impact of noise interference.

Furthermore, we observe a scarcity of publicly accessible RGB-SAR datasets, many of which exhibit uneven quality. To address this lack, we initiated a data collection campaign that gathered RGB and SAR imagery from the Pearl River Delta region in China. The collected images were carefully positioned correction, and refined annotations were meticulously applied. Specifically, our dataset comprises regions with cloud cover, offering a challenge for evaluating the effectiveness of RGB-SAR fusion under less-than-ideal observation conditions. Unlike datasets that artificially simulate cloud and fog conditions, ours is more appropriate for intelligent interpretation tasks in real-world application scenarios.

The main contributions of this paper are as follows:

1) We introduce ASANet, which is distinguished to effectively calibrate and align the information of the two modalities by its ability to leverage the complementarity and consistency of feature information across different modalities.

2) We have constructed a dataset with fine-grained labels, named PIE-RGB-SAR, which includes RGB and corresponding SAR images. The land categories are organized into six distinct classes.

3) Our proposed ASANet achieves state-of-the-art (SOTA) performance on the three datasets. Specifically, on the PIE-RGB-SAR dataset, our ASANet outperforms other methods, with an improvement in mean Intersection over Union (mIoU) ranging from 1.21\% to 17.69\%.

~ 

\section{Related work}\label{sec:Related work}

\subsection{Land cover classification}\label{sec:Land cover classification}

The significance of LCC in remote sensing images cannot be overstated, given its pivotal role in addressing a wide range of practical application requirements \citep{li2022deep}. Within this context, semantic segmentation is extensively employed to guarantee accurate classification of various land cover categories. Initially, researchers relied on machine learning techniques, such as random forest \citep{rodriguez2012assessment} and K-means clustering \citep{he2014enhanced}, to segment high-resolution remote sensing images, which aided in the efficient extraction of surface features.

With the rapid advancement of deep learning techniques, models based on Convolutional Neural Networks \citep{yang2018classification} and the Transformer \citep{zhou2023building} series have become widely utilized in remote sensing image analysis. For instance, SLCNet \citep{yu2023long}, which integrates global and local feature correlations, demonstrates superior performance across three remote sensing datasets. Nevertheless, fully supervised semantic segmentation methods \citep{costa2018supervised,zhao2023land} are increasingly challenged in meeting the demands of large-scale land mapping tasks. \citep{ma2023domain} introduced an alternative approach in which an unsupervised domain adaptation method for LCC is developed, leveraging local consistency and global diversity. This method utilizes a subset of annotated source domain data along with unlabeled target domain data to improve the generalization capability of the semantic segmentation model. This adaptation strategy enhances the model’s adaptability for LCC in diverse geographical areas. Additionally, a prominent remote sensing big-model \citep{osco2023segment} aims to promote the extensive application of semantic segmentation models in real-world scenarios. While these strategies contribute to making the segmentation process more intelligent, challenges persist when dealing with RGB images captured under challenging conditions and extreme weather. Such conditions can pose difficulties for models to learn effectively and segment feature classes accurately.

SAR has garnered considerable attention from researchers due to its microwave-active imaging mechanism, which enables all-weather imaging capabilities. Early studies \citep{orlikova2019land} exploited the sensitivity of different polarization SAR images to feature types for LCC using machine learning. Additionally, the segmentation of water and soil areas \citep{guo2022water} could be relatively easily achieved by utilizing the water path scattering property of SAR. However, the coherent imaging mechanism of SAR introduces a high degree of variability at the individual pixel level, known as coherent scattering \citep{luchaoran2021}, which poses significant challenges for LCC. The GLNS \citep{liu2022high} addresses this challenge by integrating both local and global features to extract more efficient and discriminative features and addresses the issue of excessive inter-class distances in SAR by incorporating a twofold loss function. \citep{liu2019statistical} proposed combining representation learning with statistical analysis to capture the statistical properties of SAR in feature space. While these methods offer various perspectives for mitigating the interference caused by coherent scattering in SAR, the LCC task based on SAR remains challenging. This challenge is primarily due to the lack of rich texture and the absence of color information in SAR images.

~ 

\subsection{Multimodal segmentation algorithm}\label{sec:Multimodal segmentation algorithm}

In challenging real-world application scenarios, single modal algorithms often struggle to fully meet the task requirements. For example, in tasks like semantic segmentation for autonomous driving in foggy conditions \citep{rizzoli2022multimodal}, intelligent interpretation of remote sensing images with cloud cover \citep{li2023aligning}, or medical image diagnosis relying on multimodal data \citep{hermessi2021multimodal}, the complexities of these real-world scenarios have led to the development of multimodal segmentation algorithms. 

As shown in Fig. \ref{fig:multimodel}a), the model trains on the fused images generated from the data of the two modalities to obtain segmentation results, or it fuses the segmentation results of the two modalities and calculates the loss. \citep{xu2023multi} proposed generating the RGB cloud-removed image from the RGB-SAR dataset, which is subsequently used for LCC. Some scholars have also employed methods such as Principal Component Analysis, HSV Color Space, and Generative Adversarial Networks \citep{bermudez2019synthesis} to create fused images, which are then input into the network for training. The quality of the generated images significantly affects the segmentation results \citep{he2018multi}. \citep{hang2020classification} introduced a network where the post-processing results are obtained through the weighted summation of the segmentation results from the two modalities. The loss is calculated by comparing these results with the ground truth, providing feedback for network learning. However, this method encounters challenges in learning the complementary relationship between the two modalities. To address these issues, MMFNet \citep{li2023progressive} computes phase features of RGB and SAR images as intermediate features for multi-level training, thereby avoiding direct feature interaction between the modalities. Nonetheless, the multi-level network model presents a complex structure requiring longer training time.

Typically, symmetric Siamese networks are employed to construct the multimodal framework, which facilitates the adaptive learning of feature correlations between the two modalities. As depicted in Fig. \ref{fig:multimodel}b), feature fusion occurs within the encoder or decoder section of the framework, and one or more segmentation heads are deployed to generate the final segmentation outcome. Methods such as FuseNet \citep{hazirbas2017fusenet} and the network proposed by \citep{gao2023distilled} fuse feature maps from both modalities at the decoder stage and then feed them into the segmentation head for the segmentation process. CMFNet \citep{ma2022crossmodal} and FTransUNet \citep{ma2024multilevel} propose the use of cross-scale fusion modules, which are designed to integrate features across different modalities and at multiple levels. This fusion strategy integrates the feature extraction component, facilitating the learning of feature correlations. However, it does not provide a feedback mechanism to the branches for feature interaction. The attention weights for feature interactions within SwinTFNet are adjusted exclusively for the RGB branches. CMX \citep{zhang2023cmx} introduces an attention module named CM-FRM, which is specifically designed to learn the weights of two modal channel and spatial features through feature interaction. These interacted feature maps are then passed to the subsequent layer for further learning. The CM-FRM module enhances the network’s capability to integrate feature information from another modality within the encoder. Other networks, including ACNet \citep{hu2019acnet}, SA-Gate \citep{chen2020bi}, and CroFuseNet \citep{wu2023crofusenet}, also utilize distinct attention modules in their encoder sections to calculate feature weights between the two modalities. These weights are employed to selectively fuse features across the two modalities. While these interactive modules calculate feature weights that highlight the correlation between features of the two modalities and underscore the significance of feature fusion, they often fail to consider the feature independence within each modality. These interaction modules highlight the correlation between features of two modalities and emphasize the importance of feature fusion. However, they often ignore the feature independence within the RGB and SAR modalities, and only output a feature weight that leads to the same feature representation for both modalities to learn.

\section{Method}\label{sec:Method}

In this section, we present a comprehensive overview of our proposed Asymmetric Semantic Aligning Network, which comprises the Semantic Focusing Module and the Cascade Fusion Module.

\begin{figure*}
    \centering
    \includegraphics[width=\linewidth]{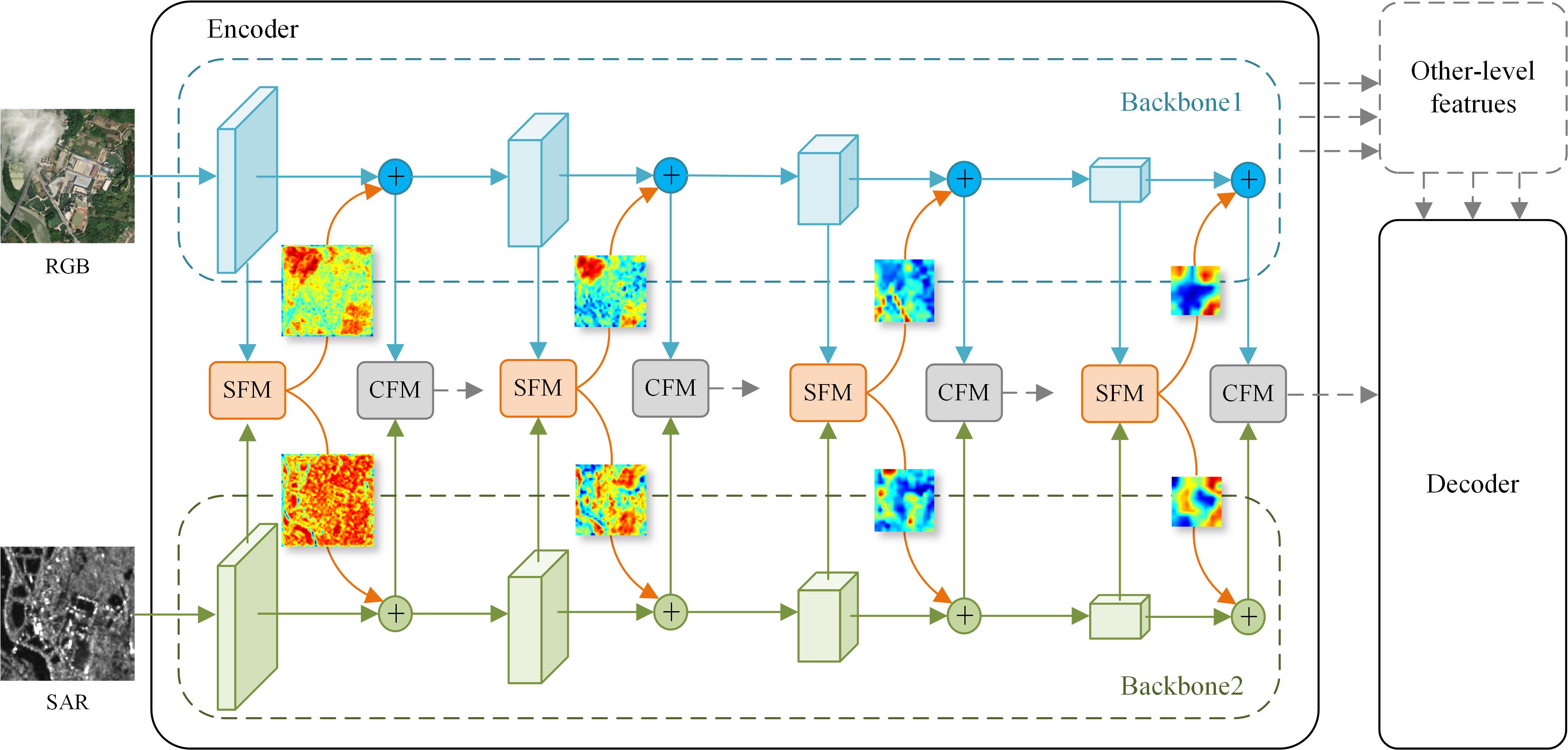}
    \caption{Overall Network Structure of ASANet: The SFM concentrates on independently complementary features of different modalities, and the CFM calibrates and aligns feature information. ($\oplus$ denotes the pixel-wise add operation)}
    \label{fig:ASANet}
\end{figure*}

\subsection{Overall network architecture}\label{sec:Overall network architecture}

The overall network structure of ASANet is provided in Fig. \ref{fig:ASANet}, which is structured on a dual-branch encoder and decoder architecture. In the encoder, we employ two branches where weights are not shared to concurrently extract features from RGB and SAR images, thereby capturing the unique characteristics of each modality in a parallel and interactive manner. The model’s residual structure enables it to discern and leverage the substantial differences among various data modalities, resulting in the generation of more effective and complementary features for subsequent stages of feature extraction and interaction. To optimally fuse the two modal features after focusing on the SFM, we propose the CFM. The decoder can be based on nearly any design from SOTA segmentation networks. For broad applicability, we select UPerNet \citep{xiao2018unified} as our preferred decoder design.

\subsection{Semantic Focusing Module}\label{sec:SFM}

\begin{figure}[h]
    \centering
    \includegraphics[width=\linewidth]{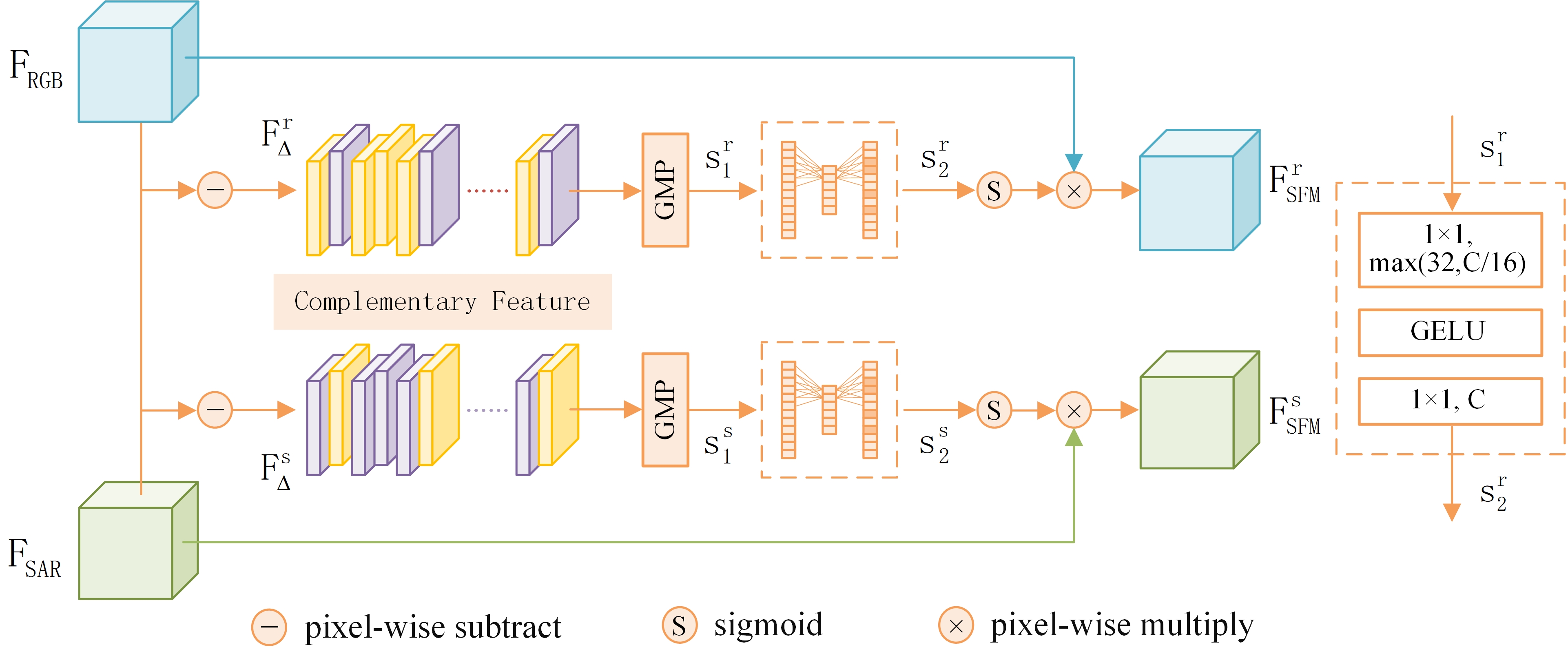}
    \caption{Diagram of the structure of the SFM.}
    \label{fig:sfm}
\end{figure}

Based on the analysis above, it is clear that the feature information required for RGB and SAR images is usually different. Consequently, this study introduces an innovative SFM that obtains different information to adjust the features of parallel branches. Drawing from the channel attention mechanism \citep{hu2018squeeze}, the SFM handles multimodal features along the channel axis. Unlike previous methods which use a single attention matrix to weigh different branches, our approach capitalizes on both consistent and complementary features. The architecture of the SFM is shown in Fig. \ref{fig:sfm}. Its design allows the SFM to compute attention matrices for individual branches, focusing on distinctive and complementary features between them. This provides a robust foundation for subsequent network layers to accurately calibrate and align features from the two modalities.

Specifically, the process comprises three main steps. We denote the feature maps of the two modalities as $\mathrm{F_{RGB}\in\mathbb{R}^{C\times H\times W}}$ and $\mathrm{F_{SAR}\in\mathbb{R}^{C\times H\times W}}$, respectively, where C, H, and W correspond to the number of channels, height, and width of the feature maps. Superscripts r (RGB) and s (SAR) are utilized to distinguish between the feature maps of the two modalities.

\textbf{Obtain differential features}: As previously mentioned, to capture the complementarity between different modalities, we extract complementary features from each modality separately. Following the principle of differential amplification \citep{qingyun2022cross}, the SFM performs a pixel-wise subtraction between the $\mathrm{F_{RGB}}$ and $\mathrm{F_{SAR}}$ feature maps to obtain the differential feature map $\mathrm{F_\bigtriangleup ^r\in\mathbb{R}^{C\times H\times W}}$ for the RGB modality. Conversely, for the SAR modality, a subtraction is performed between $\mathrm{F_{RGB}}$ and $\mathrm{F_{SAR}}$ to obtain the differential feature map $\mathrm{F_\bigtriangleup ^s\in\mathbb{R}^{C\times H\times W}}$. The first step can be described as Eq. \ref{equ-1}:

\begin{equation} \label{equ-1} 
\begin{aligned}
&\mathrm{F_\bigtriangleup ^r=F_{RGB}-F_{SAR}}\\
&\mathrm{F_\bigtriangleup ^s=F_{SAR}-F_{RGB}}\\
\end{aligned}
\end{equation}

where - denotes the pixel-wise subtraction operation.

\textbf{Adaptive refinement learning:} Furthermore, we apply global max-pooling to the differential feature maps $\mathrm{F_\bigtriangleup ^r}$ and $\mathrm{F_\bigtriangleup ^s}$ to obtain channel weights $\mathrm{s_1^r\in\mathbb{R}^{C\times 1\times 1}}$ and $\mathrm{s_1^s\in\mathbb{R}^{C\times 1\times 1}}$. Subsequently, we introduce multiple convolutional modules resembling channel-wise perceptrons to perform fine-grained learning on $\mathrm{s_1^r}$ and $\mathrm{s_1^s}$, resulting in global differential weights $\mathrm{s_2^r\in\mathbb{R}^{C\times 1\times 1}}$ and $\mathrm{s_2^s\in\mathbb{R}^{C\times 1\times 1}}$. Eq. \ref{equ-2} describes the second step.

\begin{equation} \label{equ-2}
    \begin{aligned}
        &\mathrm{s_1^r=\mathcal {F}_{GMP}(F_\bigtriangleup ^r)} \\
        &\mathrm{s_2^r=\mathcal{F}_2(\mathcal{F}_{GELU}(\mathcal{F}_1(s_1^r)))}\\
        &\mathrm{s_1^s=\mathcal {F}_{GMP}(F_\bigtriangleup ^s)} \\
        &\mathrm{s_2^s=\mathcal{F}_2(\mathcal{F}_{GELU}(\mathcal{F}_1(s_1^s)))} \\
    \end{aligned}
\end{equation}

where $\mathrm{\mathcal{F}_{GMP}}$ denotes global max-pooling, $\mathrm{\mathcal{F}_1}$ represents a convolutional layer with an output dimension of $\mathrm{max(32,\frac C{16})}$, $\mathrm{\mathcal{F}_2}$ is a convolutional layer with an output dimension of C, and $\mathrm{\mathcal{F}_{GELU}}$ signifies the GELU activation function.

\textbf{Correction of parallel branches:} The formula for correcting the original feature maps is as follows:

\begin{equation} \label{equ-3}
    \begin{aligned}
        &\mathrm{F_{SFM}^r=F_{RGB}\otimes \sigma (s_2^r)} \\
        &\mathrm{F_{SFM}^s=F_{SAR}\otimes \sigma (s_2^S)} 
\end{aligned}
\end{equation}

where $\mathrm{F_{SFM}^r\in\mathbb{R}^{C\times H\times W}}$ and $\mathrm{F_{SFM}^s\in\mathbb{R}^{C\times H\times W}}$ represent the output feature maps of the dual branches, respectively. $\mathrm{\sigma}$ is the sigmoid function and $\mathrm{\otimes}$ is the pixel-wise multiplication.

\subsection{Cascade Fusion Module}\label{sec:CFM}

\begin{figure}[h]
    \centering
    \includegraphics[width=\linewidth]{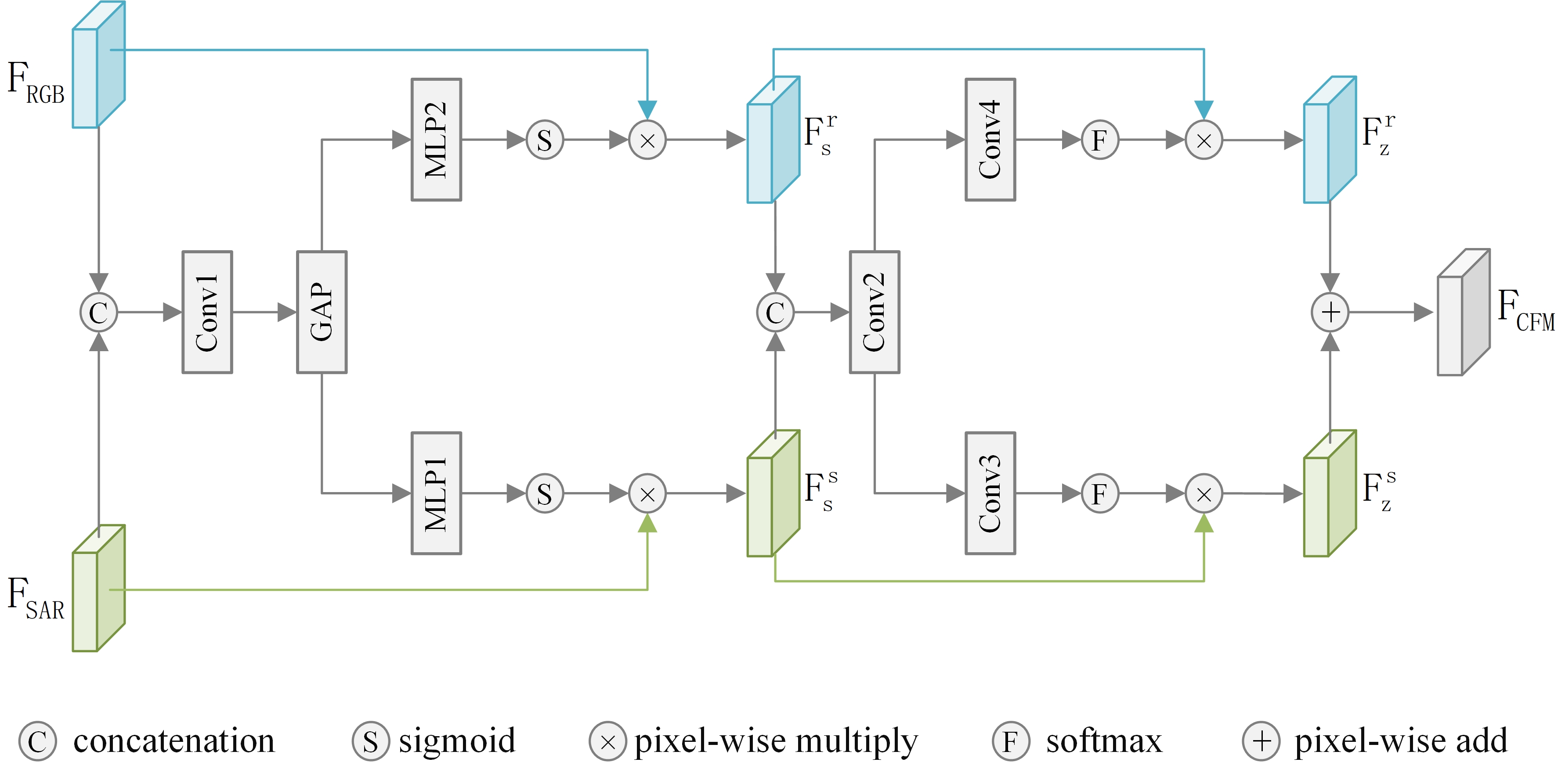}
    \caption{Diagram of the structure of the CFM.}
    \label{fig:cfm}
\end{figure}

The feature information in RGB and SAR images is often complementary. The challenge lies in effectively calibrating and aligning the complementary information from both modalities. Traditional feature fusion strategies, such as simple concatenation or summation of feature maps, may inadvertently amplify the noise present in both modalities, thereby negatively impacting the final classification performance. To overcome this issue, inspired by CBAM \citep{woo2018cbam}, we propose a CFM for feature fusion. As illustrated in Fig. \ref{fig:cfm}, the CFM adopts a concatenated attention mechanism to learn deep representations of RGB and SAR features in both channel and spatial dimensions. This fusion approach can adaptively calibrate and align the complementary information from the two modalities, reducing the interference of noise information and improving the model’s performance.

Suppose the input feature maps to the CFM are $\mathrm{F_{RGB}\in\mathbb{R}^{C\times H\times W}}$ and $\mathrm{F_{SAR}\in\mathbb{R}^{C\times H\times W}}$, where C, H, and W represent the number of channels, height, and width of the feature maps, respectively. The superscripts r (RGB) and s (SAR) are used to distinguish between the feature maps of the two modalities.

In the first attention module, we align and calibrate the features from the two branches along the channel dimension. Initially, the feature maps $\mathrm{F_{RGB}}$ and $\mathrm{F_{SAR}}$ are concatenated. Subsequently, the Concatenated feature map is fused through convolutional layers, followed by global average-pooling to obtain the feature vector $\mathrm{s \in \mathbb{R}^{C \times 1 \times 1}}$. GAP generate channel-wise statistics by shrinking concatenated feature through its spatial dimensions. The feature vector s is then input into two separate Multi-Layer Perceptrons (MLPs) with non-shared weights to capture channel-wise dependencies. After passing through the sigmoid function, different fusion channel weights are obtained for each modality. These weights are then multiplied with the original feature maps to obtain the feature maps $\mathrm{F_s^r \in \mathbb{R}^{C \times H \times W}}$ and $\mathrm{F_s^s \in \mathbb{R}^{C \times H \times W}}$. The process is described in formulaic terms as follows:

\begin{equation}\label{equ-4}
\begin{aligned}
&\mathrm{s=\mathcal {F}_{GAP}(\mathcal {F}_1^c(\mathit{cat}(F_{RGB},F_{SAR})))} \\
&\mathrm{F_s^r=F_{RGB}\otimes \sigma (\mathcal {F}_{MLP_1}(s))} \\
&\mathrm{F_s^s=F_{SAR}\otimes \sigma (\mathcal {F}_{MLP_2}(s))} \\
\end{aligned}
\end{equation}

where $\mathit{cat}$ refers to concatenating feature maps along the channel dimension, $\mathrm{\mathcal {F}_{GAP}}$ denotes global average-pooling, $\mathrm{\mathcal{F}_1^c}$ represents a convolutional layer with an input dimension of 2C and an output dimension of c. $\mathrm{\mathcal{F}_{MLP_1}}$ and $\mathrm{\mathcal{F}_{MLP_2}}$ signify MLP. $\mathrm{\sigma}$ represents the sigmoid activation function, $\mathrm{\otimes}$ denotes the pixel-wise multiplication operation. CFM learn different channel attention weights, enabling the adaptive selection and calibration of the original features from the two branches.

In the second attention module, we align and calibrate the features from the two branches along the spatial dimensions. Initially, the feature maps $\mathrm{F_s^r}$ and $\mathrm{F_s^s}$ are concatenated, and the concatenated feature map undergoes processing through a convolutional layer to fuse the information, yielding feature map z. This feature map $\mathrm{z \in \mathbb{R}^{1 \times H \times W}}$ is then convolved with two sets of distinct weights to enhance the feature representation. The softmax function is applied to derive the weight scores for the respective modal space dimensions, which represent the importance of each pixel. These scores are used to weight the feature maps $\mathrm{F_s^r}$ and $\mathrm{F_s^s}$, resulting in the fusion spatial attention weighted feature maps $\mathrm{F_z^r}$ and $\mathrm{F_z^s}$.

\begin{equation}\label{equ-5}
\begin{aligned}
&\mathrm{z=\mathcal {F}_2^c(\mathit{cat}(F_s^r,F_s^s))} \\
&\mathrm{F_z^r=F_s^r\otimes \mathcal{F}_{softmax}(\mathcal{F}_3(z))} \\
&\mathrm{F_z^s=F_s^s\otimes \mathcal{F}_{softmax}(\mathcal{F}_4(z))} \\
\end{aligned}
\end{equation}

where $\mathit{cat}$ refers to concatenating feature maps along the channel dimension, $\mathrm{\mathcal{F}_2^c}$ represents a convolutional layer with an input dimension of 2C and an output dimension of c, $\mathrm{\mathcal{F}_{3}}$ and $\mathrm{\mathcal{F}_{4}}$ represent convolutional layers with both input and output dimensions C. $\mathrm{\otimes}$ represents pixel-by-pixel multiplication, and $\mathrm{\mathcal{F}_{softmax}}$ is a softmax function. CFM calculate the spatial attention weights for different modalities, adaptively strengthening the effective features and suppressing the invalid ones.

Finally, According to Eq. \ref{equ-6}, $\mathrm{F_z^r}$ and $\mathrm{F_z^s}$ are pixel-wise added to produce the composite output feature map $\mathrm{F_{CFM}\in \mathbb{R}^{C \times H \times W}}$.

\begin{equation}\label{equ-6}
\begin{aligned}
&\mathrm{F_{CFM}=F_z^r+F_z^s}
\end{aligned}
\end{equation}
where + denotes the pixel-wise addition (PWA) operation.

\section{Experimental and Datasets}\label{sec:Experimental and datasets}

This section describes the proposed refined annotated semantic segmentation multimodal dataset. Furthermore, a comparative analysis is performed on this dataset, in conjunction with two publicly available datasets, utilizing six other multimodal segmentation algorithms. This comparison aims to highlight the superior performance of the proposed algorithm. The experimental results consistently demonstrate the robustness of our algorithm, affirming its efficacy in handling diverse multimodal datasets.

\subsection{Datasets}\label{sec:Datasets}

\subsubsection{PIE-RGB-SAR}\label{sec:PIE-RGB-SAR}

\begin{figure}[h]
    \centering
    \includegraphics[width=\linewidth]{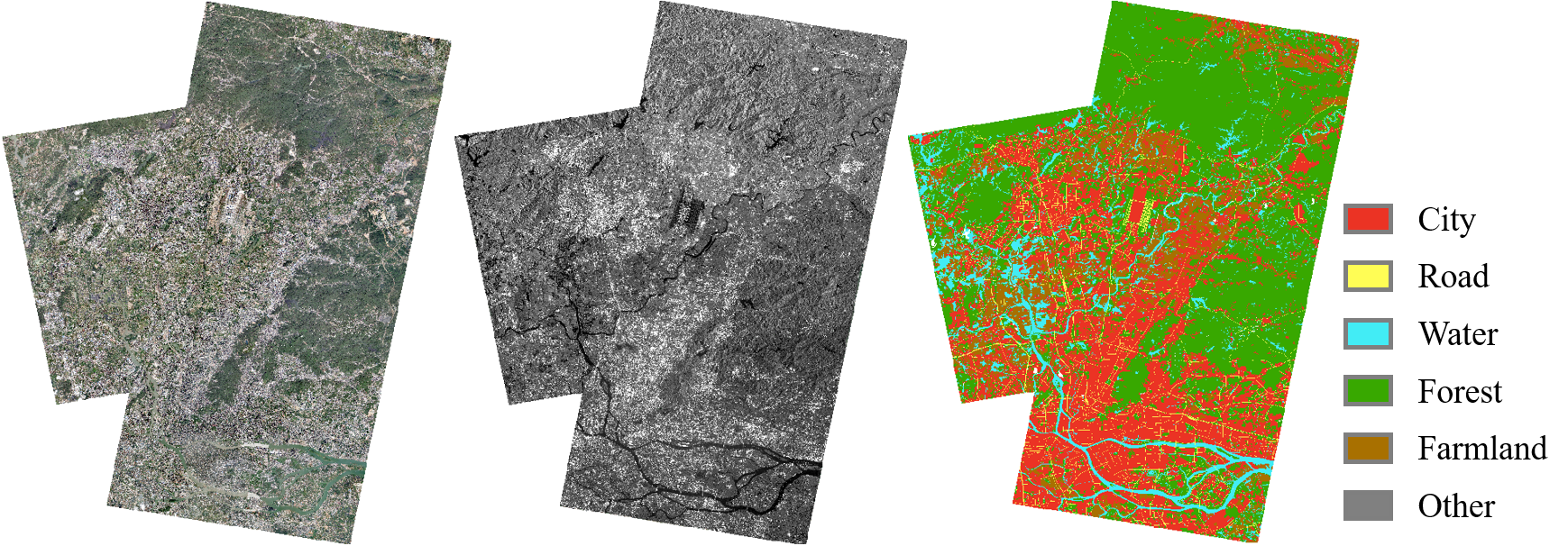}
    \caption{PIE-RGB-SAR dataset: RGB image on the left, SAR image in the middle, and ground truth image on the right.}
    \label{fig:pie-dataset}
\end{figure}

As shown in Fig. \ref{fig:pie-dataset}, the study area for this paper is selected as a part of the Pearl River Delta, China. The RGB image is sourced from Google satellite data with a spatial resolution of approximately 0.5 meters. The SAR image is derived from the GF3 satellite’s ultra-fine stripe mode, with a spatial resolution of 3 meters. To ensure the data resolution of the two modes remains consistent, we utilize PIE-Basic software for resampling and aligning the two images. To reduce interference from extreme radiation intensity during the network’s learning from SAR images and to facilitate network convergence, radiation values in SAR images are truncated and linearly adjusted to range between 0 and 255.

Considering that the RGB and SAR images were not captured simultaneously, there are regions where feature changes have occurred across different categories. In this dataset, the RGB image serves as the primary source, with the SAR image offering complementary information to aid in identifying the land type associated with each pixel. The dataset is annotated into six categories: city, road, water, forest, farmland, and others. It is noteworthy that the dataset exhibits an imbalance in data distribution, with distinct proportions for each feature type. Specifically, the distribution consists of 30\% for city, 3\% for road, 7\% for water, 42\% for forest, 10\% for farmland, and 8\% for other categories. The images were cropped to a uniform size of 256x256, yielding a total of 4,865 images. These images were subsequently partitioned into training and validation sets at a 1:1 ratio, comprising 2,433 images in the training set and 2,432 images in the validation set.

\subsubsection{DDHR-SK}\label{sec:DDHR-SK}

\begin{figure}[h]
    \centering
    \includegraphics[width=\linewidth]{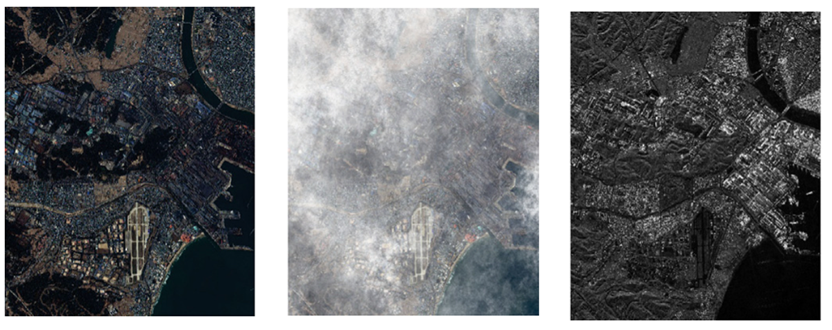}
    \caption{DDHR-SK dataset: Original RGB image on the left, RGB image with simulated cloudy in the center, and SAR image on the right.}
    \label{fig:ddhr-sk-dataset}
\end{figure}

\citep{ren2022dual} provided several multimodal datasets, which are divided into five image groups and sourced from three distinct regions. Our experiments were conducted using data from the cloudy group in the Pohang region of South Korea, known as DDHR-SK, where the data division is publicly available. The dataset encompasses five categories: city, road, forest, water, and farmland. The overall preview of the DDHR-SK dataset is shown in Fig. \ref{fig:ddhr-sk-dataset}. The origin RGB image is sourced from GF2 satellite data. The RGB images of the DDHR-SK dataset were processed using the GNU Image Manipulation Program (GIMP) to simulate a cloudy scene. The SAR image is sourced from GF3 satellite data and has been adjusted to have a resolution of 1 meter. The RGB and SAR images are cropped to a pixel size of 256x256. There are 3087 images for training and 3086 images for verification. This dataset aims to capture the challenges posed by foggy and cloudy conditions and serves as a valuable resource for evaluating multimodal segmentation algorithms.

\subsubsection{WHU-OPT-SAR}\label{sec:WHU-OPT-SAR}

\begin{figure}[h]
    \centering
    \includegraphics[width=\linewidth]{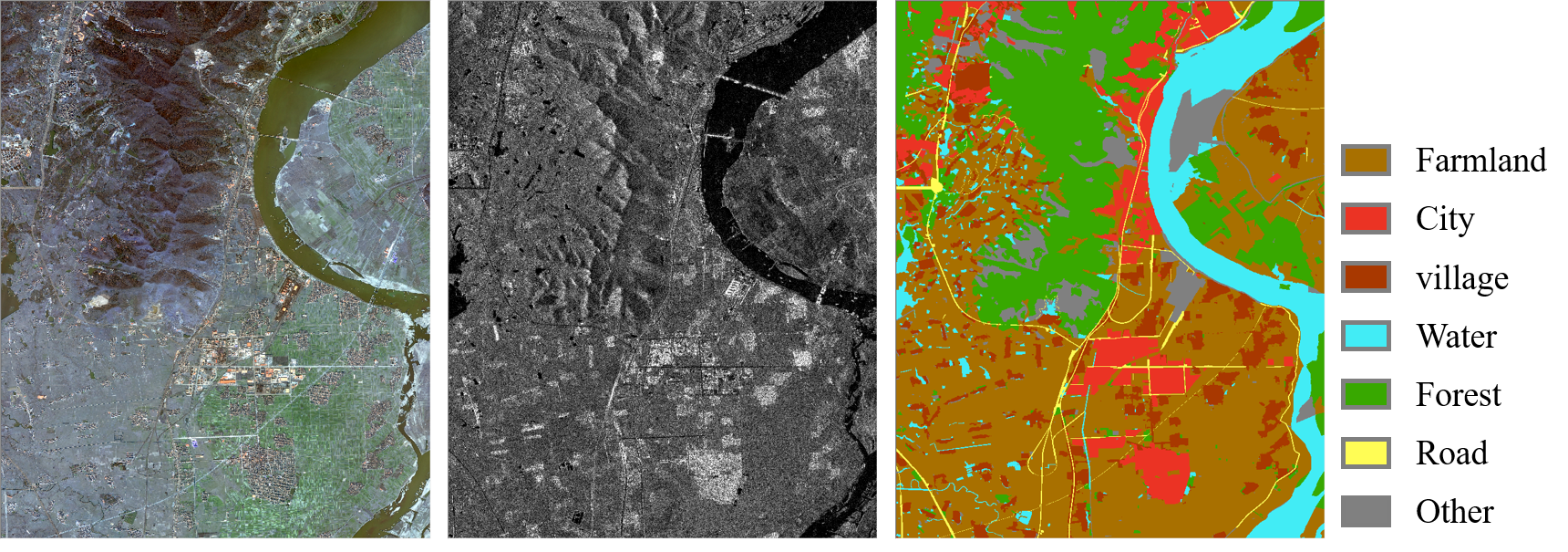}
    \caption{Part of the WHU-OPT-SAR dataset: RGB image on the left, SAR image in the center, and ground truth image on the right. }
    \label{fig:whu-dataset}
\end{figure}

The WHU-OPT-SAR dataset \citep{li2022mcanet} is a multimodal dataset that features a study area in Hubei, China. The dataset comprises a total of 100 RGB and SAR images, each with dimensions of 5,536x3,704 pixels. The RGB images are sourced from GF1 satellite data, while the SAR images are obtained from GF3 satellite data, with a resolution of 5 meters. One of the dataset images is shown in Fig. \ref{fig:whu-dataset}. In this paper, the images are cropped to a size of 512x512 pixels, resulting in a total of 7,040 images for training and 1,760 images for validation. The dataset is annotated into seven categories: farmland, city, village, water, forest, road, and other. This dataset provides a valuable resource for evaluating multimodal segmentation algorithms, particularly in the context of diverse land cover types in Hubei, China.

\subsection{Experimental setups}\label{sec:Experimental setups}

In this paper, all model training and testing are conducted on two Quadro RTX 8000 GPUs, which are running on the Ubuntu 18.04 Linux platform. The batch size per GPU is set to 4, and a total of 80,000 iterations are performed. The model is validated every 8000 iterations. The metrics reported in this paper are indicators of the number of iterations corresponding to the best mIoU. The results for all models are based on the accuracy of our own reproduction. The hyperparameter settings for each model are listed in the table. Specifically, pre-trained weights from ImageNet22k are adopted to initialize two ConvNeXtV2-tiny \citep{woo2023convnext} backbone networks within ASANet. The primary data augmentation techniques include random scaling, random cropping, and random flipping. To ensure the fairness of the experiment, all experimental configurations outlined in this paper were strictly adhered to throughout the study. The evaluation metrics include mIoU, Kappa, and Overall Accuracy (OA), which are commonly employed in semantic segmentation assessments. We also used floating point operation count (FLOPs), the number of model parameters, and FPS to evaluate the complexity of the model (see Table \ref{tab:setting}).

\begin{table}[!ht]
    \centering
    \caption{Settings of hyperparameters for each model.}
    \resizebox{1.0\linewidth}{!}{
    \begin{tabular}{|c|c|c|c|c|c|c|}
    \hline
        \textbf{Model} & \textbf{Batchsize} & \textbf{Optimizer} & \textbf{Ir} & \textbf{Momentum/Betas} & \textbf{Weight-decay} \\ \hline
        UPerNet & \multirow{8}{*}{8} & AdamW & 0.0001 & (0.9,0.999) & 0.05 \\ \cline{1-1} \cline{3-6}
        FuseNet &  & SGD & 0.01 & 0.9 & 0.0005 \\ \cline{1-1} \cline{3-6}
        SA-Gate &  & SGD & 0.01 & 0.9 & 0.0005 \\ \cline{1-1} \cline{3-6}
        AFNet &  & SGD & 0.01 & 0.9 & 0.0005 \\  \cline{1-1} \cline{3-6}
        CMFNet &  & SGD & 0.01 & 0.9 & 0.0005 \\  \cline{1-1} \cline{3-6}
        CMX &  & AdamW & 0.00006 & (0.9,0.999) & 0.01 \\ \cline{1-1} \cline{3-6}
        FTransUNet &  & SGD & 0.01 & 0.9 & 0.0005 \\  \cline{1-1} \cline{3-6}
        ASANet(ours) &  & AdamW & 0.0001 & (0.9,0.999) & 0.05 \\ \hline
    \end{tabular}
    }
\label{tab:setting}
\end{table}

\begin{table}[h]
    \centering
    \caption{The complexity computation metrics for a 256 × 256 image on a single Quadro RTX 8000 GPU are as follows. The mIoU metrics (\%) for different methods across three datasets are presented below. The best results are presented in bold, while the second-best results are underlined.}
    \resizebox{1.0\linewidth}{!}{
    \begin{tabular}{cccccccc}
    \hline
        \multirow{2}{*}{\textbf{Model}} & \multirow{2}{*}{\textbf{Year}} & \textbf{FLOPs} & \textbf{Parameter} & \textbf{Speed} & \multicolumn{3}{c}{\textbf{mIoU}} \\ 
         & & \textbf{G} & \textbf{M} & \textbf{FPS} & \textbf{PIE-RGB-SAR} & \textbf{DDHR-SK} & \textbf{WHU-OPT-SAR} \\ \hline
        FuseNet & 2017 & 66 & \textbf{55} & \textbf{88.8} & 60.62 & 48.87 & 38.01 \\ 
        SA-Gate & 2020 & 46 & 121 & 34.9 & 73.84 & 90.89 & 53.17 \\ 
        AFNet & 2021 & 65 & 356 & 35.9 & 76.27 & 91.11 & 53.57 \\ 
        CMFNet & 2022 & 77 & 104 & 21.6 & 76.31 & 89.79 & 53.72 \\
        CMX & 2023 & \textbf{15} & \uline{67} & 33.5 & \uline{77.10} & \uline{94.32} & \uline{55.68} \\ 
        FTransUNet & 2024 & 70 & 203 & 20.7 & 75.72 & 87.64 & 54.47 \\
        \textbf{ASANet(ours)} & & \uline{25} & 82 & \uline{48.7} & \textbf{78.31} & \textbf{94.48} & \textbf{56.11} \\ \hline
    \end{tabular}
    }
\label{tab:mIoU}
\end{table}

\subsection{Experimental}\label{sec:Experimental}

To evaluate the models’ performance, we conducted a detailed comparison of six multimodal semantic segmentation models: CMX \citep{zhang2023cmx}, SA-Gate \citep{chen2020bi}, FTransUNet \citep{ma2024multilevel}, CMFNet \citep{ma2022crossmodal}, AFNet \citep{xu2021attention}, and FuseNet \citep{hazirbas2017fusenet}. The mIoU performance of various methods on three datasets is presented in Table \ref{tab:mIoU}, where our proposed ASANet demonstrates SOTA results. ASANet inference is 15.2 FPS faster than the suboptimal model CMX. Furthermore, we tested the UperNet model, which employs the ConvNeXtV2-tiny backbone, for single-modal experiments. All models were configured to ensure comparability, facilitating a fair evaluation of both multimodal and single-modal segmentation capabilities.

\subsubsection{Results on PIE-RGB-SAR}\label{sec:Results on PIE-RGB-SAR}

In Table \ref{tab:ex_pie}, we compare the performance of single model and six multimodal algorithms. In the evaluation of single model training, the RGB achieves a mIoU that is 10.17\% higher than that of SAR, suggesting its significant contribution to feature learning. ASANet (mIoU=78.31\%, Kappa=85.27\%, OA=89.64\%) surpasses other methods in all metrics across both modalities and modules. Section \ref{sec:Analysis of ablation results for different modules} offers a detailed analysis of these results.

\begin{table*}[!h]
    \centering
    \caption{Metrics results (\%) for different methods on PIE-RGB-SAR dataset. The best results are presented in bold, while the second-best results are underlined.}
    \resizebox{1.0\linewidth}{!}{
    \begin{tabular}{cccccccccccc}\hline
        \multirow{2}{*}{\textbf{Model}} & \multicolumn{2}{c}{\textbf{Modality}} &\multirow{2}{*}{\textbf{Kappa}} & \multirow{2}{*}{\textbf{OA}} & \multirow{2}{*}{\textbf{mIoU}} & \multicolumn{6}{c}{\textbf{IoU}} \\ 
        & \textbf{RGB} & \textbf{SAR} & & & & \textbf{City} & \textbf{Road} & \textbf{Water} & \textbf{Forest} & \textbf{Farmland} & \textbf{Other}\\ \hline
        UPerNet & $\checkmark$ & &84.13 & 88.84 & 76.90 & 81.77 & 60.53 & 75.62 & 82.15 & 64.40 & 97.27 \\
        UPerNet & & $\checkmark$ &74.16& 83.82 & 66.73 & 75.57 & 33.69 & 70.61 & 75.76 & 49.04 & 95.73 \\
        FuseNet & $\checkmark$ & $\checkmark$ &70.79 & 79.89 & 60.62 & 70.56 & 34.47 & 60.41 & 71.01 & 29.93 & \uline{97.31} \\ 
        SA-Gate & $\checkmark$ & $\checkmark$ &82.24& 87.59 & 73.84 & 80.64 & 52.93 & 71.03 & 80.37 & 60.95 & 97.11 \\ 
        AFNet & $\checkmark$ & $\checkmark$ &83.95 & 88.71 & 76.27 & 81.90 & 58.30 & 75.03 & 82.10 & 64.06 & 96.34 \\ 
        CMFNet & $\checkmark$ & $\checkmark$ & 83.99 & 88.74 & 76.31 & 82.33 & 57.46 & \uline{77.28} & 82.06 & 61.76 & 96.96 \\
        CMX & $\checkmark$ & $\checkmark$ & \uline{84.60} & \uline{89.18} & \uline{77.10} & \uline{82.41} & \uline{59.19} & 75.88 & \uline{82.73} & \uline{65.22} & 97.06 \\ 
        FTransUNet & $\checkmark$ & $\checkmark$ & 83.53 & 88.38 & 75.72 & 81.86 & 56.66 & 77.01 & 81.67 & 59.91 & 97.21 \\
        \textbf{ASANet(ours)} & $\checkmark$ & $\checkmark$  &\textbf{85.27}& \textbf{89.64} & \textbf{78.31} & \textbf{82.85} & \textbf{61.80} & \textbf{77.83} & \textbf{83.27} & \textbf{66.75} & \textbf{97.32} \\ \hline
    \end{tabular}
    }
    
\label{tab:ex_pie}
\end{table*}

ASANet achieves notable improvements in mIoU over FuseNet, SA-Gate, FTransUNet, CMFNet, AFNet, and CMX, with increases of 15.65\%, 4.47\%, 2.59\%, 2\%, 2.04\%, and 1.21\%, respectively. FuseNet demonstrates overall lower performance. SA-Gate also underperforms on this dataset, attaining a mIoU of only 73.84\%. FTransUNet and CMFNet, networks specifically designed for remote sensing imagery, also perform poorly in terms of accuracy. CMX, a fusion algorithm designed for autonomous driving, achieves a score of 77.1\% on this dataset, which is lower than the scores achieved by ASANet.

Table \ref{tab:ex_pie} also presents the accuracy of different methods in terms of IoU metrics for individual categories. The ASANet model consistently attains the highest IoU values across all categories. Compared to the SA-Gate network, ASANet shows an 8.87\% higher $\mathrm{IoU_{road}}$ and a 6.8\% higher $\mathrm{IoU_{water}}$. Against the sub-high precision network CMX, ASANet performs better across multiple categories, with improvements of 0.44\%, 2.61\%, 1.95\%, 0.54\%, and 1.53\% for the city, road, water, forest, and farmland categories, respectively. It is worth noting that the performance of the networks FTransUNet, CMFNet, and ASANet, which are designed for remote sensing image, is outstanding in water categories. ASANet’s performance excels in the road and water categories.

\begin{figure*}[!h]
    \centering
    \includegraphics[width=\linewidth]{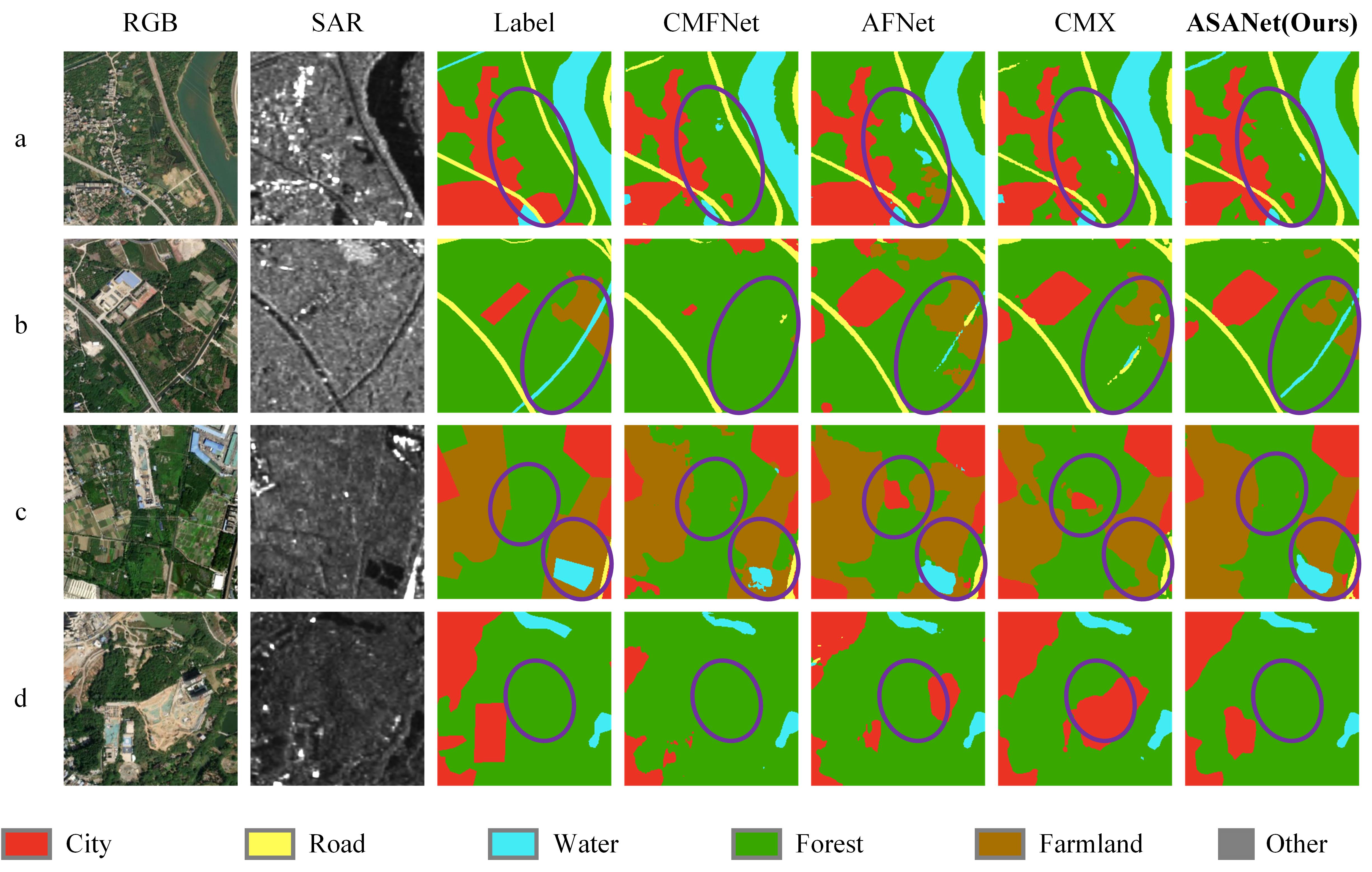}
    \caption{Visualization results of different methods on PIE-RGB-SAR dataset: four representative multimodal models.}
    \label{fig:vis_ex_pie}
\end{figure*}

The visualization results of each model on the PIE-RGB-SAR dataset are shown in Fig. \ref{fig:vis_ex_pie}. The sequence, from left to right, includes the RGB image, SAR image, ground truth label, CMFNet result, AFNet result, CMX result, and ASANet result. This visualization results compellingly evidence the superior performance of ASANet. Notably, our model exhibits improved continuity in identifying road and water features. In Fig. \ref{fig:vis_ex_pie}, groups a, b, and c show that AFNet algorithm is susceptible to SAR images, leading to confusion between forest and farmland. Conversely, ASANet demonstrates resistance to noise in SAR images. Groups c and d of Fig. \ref{fig:vis_ex_pie} highlight ASANet’s effective use of SAR information, addressing issues of omission and misjudgment present in RGB images. The visualization results underscore that ASANet skillfully integrates information from both modalities, selectively incorporating more effective data for learning.

\begin{figure*}[!h]
    \centering
    \subfigure[Label-RGB]{\includegraphics[scale=0.28]{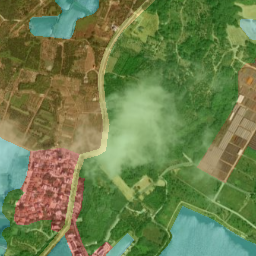}}
    \subfigure[Single-RGB]{\includegraphics[scale=0.28]{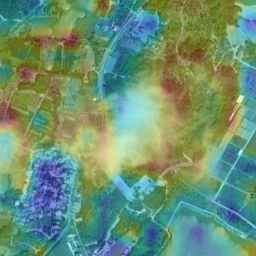}}
    \subfigure[CMX-RGB]{\includegraphics[scale=0.28]{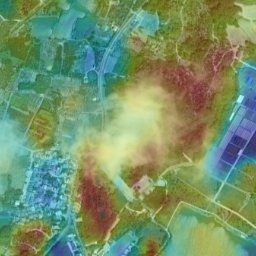}}
    \subfigure[Our-RGB]{\includegraphics[scale=0.28]{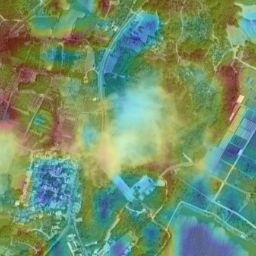}}
    \subfigure[Our]{\includegraphics[scale=0.28]{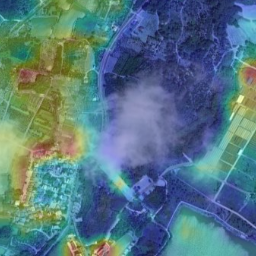}}
    \subfigure[Label-SAR]{\includegraphics[scale=0.28]{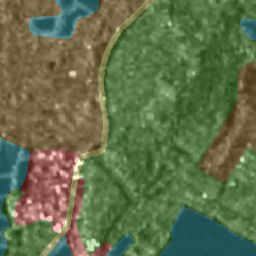}}
    \subfigure[Single-SAR]{\includegraphics[scale=0.28]{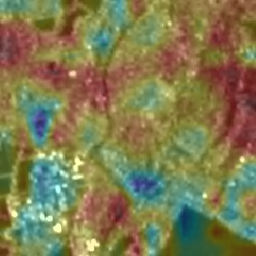}}
    \subfigure[CMX-SAR]{\includegraphics[scale=0.28]{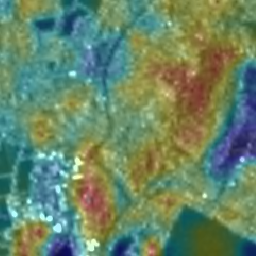}}
    \subfigure[Our-SAR]{\includegraphics[scale=0.28]{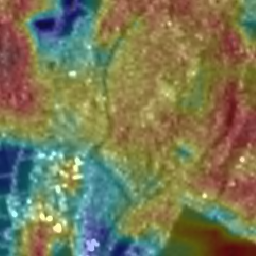}}
    \subfigure[CMX]{\includegraphics[scale=0.28]{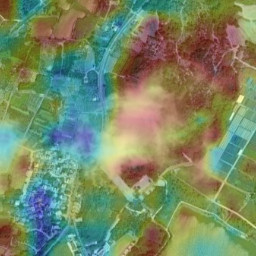}}
    \caption{Feature map visualization: (a) and (f) show the superposition ground truth of RGB and SAR images, respectively; (b) and (g) depict the output of the stage-3 backbone network of the RGB and SAR single-mode models, respectively; (c) and (h) represent the output feature map of the RGB and SAR branches after the stage-3 feature interaction of CMX (i.e., the output feature map of CM-FRM). (d) and (i) illustrate the RGB and SAR branch output feature map after the stage-3 feature interaction of our ASANet network (i.e., $\mathrm{F_{RGB}}$ and $\mathrm{F_{SAR}}$ in Fig. \ref{fig:cfm}); (e) shows the output feature map $\mathrm{F_{CFM}}$ of CFM stage-3 feature fusion in our ASANet network; (j) presents the output feature map of the FFM stage-3 in CMX.}
    \label{fig:Feature map visualization}
\end{figure*}

For a more intuitive comparison, we specifically select images with cloud cover for feature map visualization. As shown in Fig. \ref{fig:Feature map visualization}, there is a clear distinction when comparing the RGB and SAR branch features after CMX’s stage-3 feature interaction (Fig. \ref{fig:Feature map visualization}(c) and (h)) with those after ASANet’s stage-3 feature interaction (Fig. \ref{fig:Feature map visualization}(d) and (i)). In the case of CMX, the feature blending results in highly similar feature inductions for different branches. Conversely, our network, which employs the SFM strategy, fosters relatively independent features for the two branches, introducing diverse feature information from each branch. When comparing the feature maps in Fig. \ref{fig:Feature map visualization}(j) after CMX’s stage-3 feature fusion and Fig. \ref{fig:Feature map visualization}(e) after ASANet’s stage-3 feature fusion, our network consistently detects regional features of the same category with reduced noise. This results in less fragmented small areas of false detection in the segmentation results of ASANet. By utilizing the fusion strategy of SFM and CFM, ASANet excels in self-adapting the learning of both modalities, effectively incorporating SAR features while mitigating noise information. This strategic approach enhances the robustness and overall segmentation performance of ASANet.

\subsubsection{Results on DDHR-SK}\label{sec:Results on DDHR-SK}

In the DDHR-SK dataset, the optical modes consist of RGB images simulated to emulate imaging under cloudy weather conditions, adding variability to the dataset. Notably, ASANet has exhibited the most effective detection performance among other models on this dataset. This demonstrates the robustness and adaptability of ASANet to challenging scenarios, including those with simulated cloud cover in optical images.

\begin{table*}[!h]
    \centering
    \caption{Metrics results (\%) for different methods on DDHR-SK dataset. The best results are presented in bold, while the second-best results are underlined.}
    \resizebox{1.0\linewidth}{!}{
    \begin{tabular}{ccccccccccc}\hline
        \multirow{2}{*}{\textbf{Model}} & \multicolumn{2}{c}{\textbf{Modality}} & \multirow{2}{*}{\textbf{Kappa}} & \multirow{2}{*}{\textbf{OA}} & \multirow{2}{*}{\textbf{mIoU}} & \multicolumn{5}{c}{\textbf{IoU}}  \\ 
        & \textbf{RGB} & \textbf{SAR} & & & & \textbf{City} & \textbf{Road} & \textbf{Forest} & \textbf{Water} & \textbf{Farmland} \\ \hline
        UPerNet & $\checkmark$ & & 95.39 & 96.74 & 91.42 & 95.09 & 78.68 & 92.22 & 99.15 & 91.99\\
        UPerNet & & $\checkmark$ & 92.40 & 94.63 & 85.72 & 92.79 & 64.70 & 88.10 & 98.30 & 84.74\\
        FuseNet & $\checkmark$ & $\checkmark$ &  60.31 & 73.62 & 48.87 & 64.00 & 1.76 & 47.80 & 89.43 & 41.35 \\ 
        SA-Gate & $\checkmark$ & $\checkmark$ & 95.28 & 96.66 & 90.89 & 95.20 & 76.65 & 92.36 & 99.09 & 91.17 \\ 
        AFNet & $\checkmark$ & $\checkmark$ & 95.27 & 96.65 & 91.11 & 95.32 & 78.91 & 92.14 & 98.76 & 90.42\\ 
        CMFNet & $\checkmark$ & $\checkmark$ & 94.52 & 96.12 & 89.79 & 94.58 & 76.07 & 90.91 & 98.7 & 88.68 \\
        CMX & $\checkmark$ & $\checkmark$ & \uline{97.09} & \uline{97.94} & \uline{94.32} & \uline{96.95} & \textbf{85.39} & \uline{95.12} & \uline{99.46} & \uline{94.69} \\ 
        FTransUNet & $\checkmark$ & $\checkmark$ & 93.12 & 95.14 & 87.64 & 93.07 & 71.67 & 88.71 & 98.2 & 86.62 \\
        \textbf{ASANet(ours)} & $\checkmark$ & $\checkmark$ & \textbf{97.22} & \textbf{98.03} & \textbf{94.48} & \textbf{97.04} & \uline{85.36} & \textbf{95.47} & \textbf{99.49} & \textbf{95.03} \\ \hline
    \end{tabular}
    }
\label{tab:ex_ddrh}
\end{table*}

Table \ref{tab:ex_ddrh} displays the incremental results for different models and IoU metrics across all classes in the DDHR-SK dataset. When considering mIoU accuracy, the models are ranked in the following order: ASANet, CMX, AFNet, SA-Gate, CMFNet, FTransUNet and FuseNet. Each network exhibits strong performance on this dataset, aided by its high resolution. Notably, it is crucial to emphasize that the simulated images in this dataset depict scenes under cloudy conditions. As a result, the performance gap between RGB and SAR modes is substantially narrowed, suggesting that the network can still effectively learn feature information from RGB imagery despite the challenging cloud cover.

\begin{figure*}
    \centering
    \includegraphics[width=\linewidth]{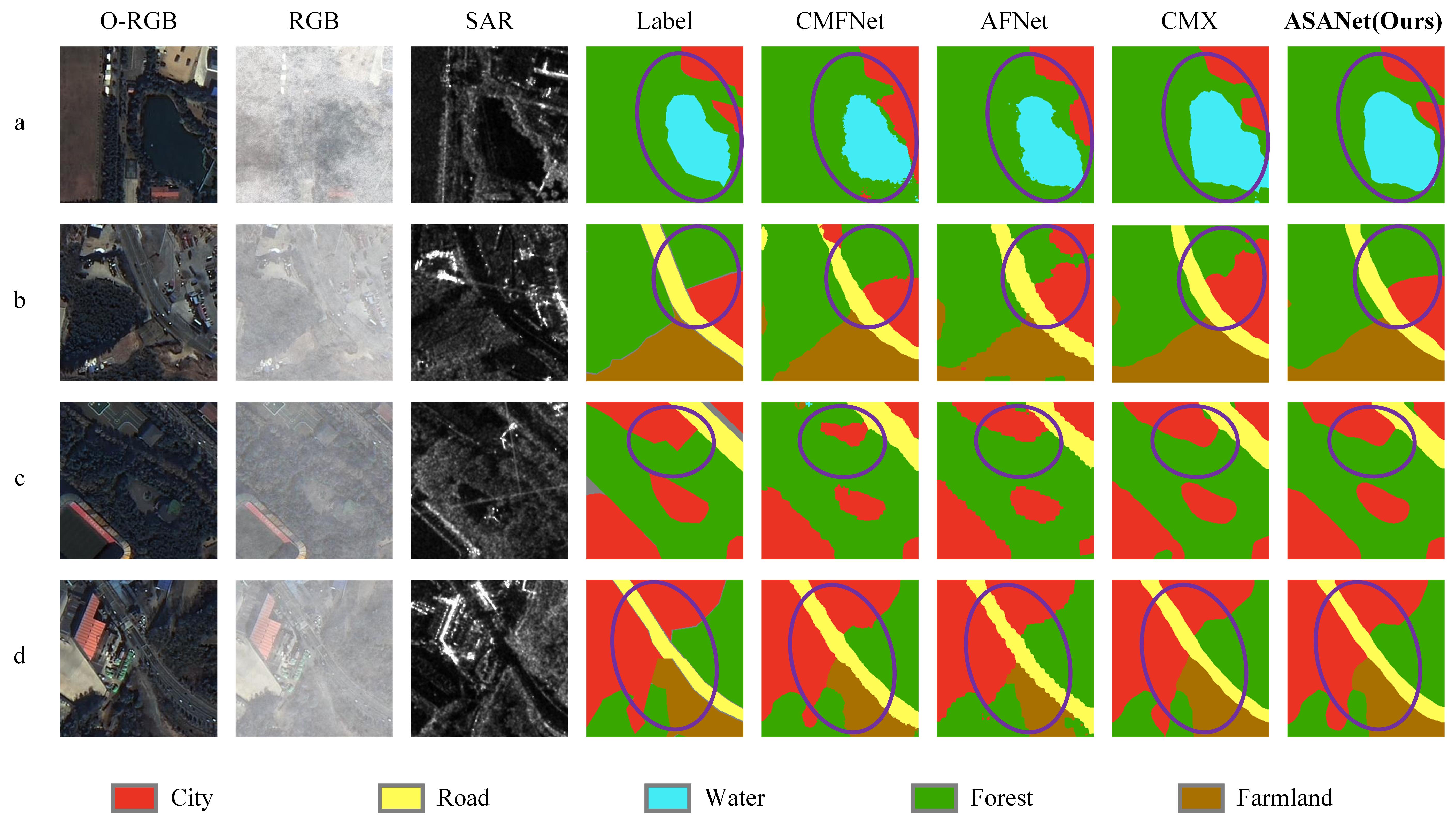}
    \caption{Visualization results of different methods on DDHR-SK dataset: four representative multimodal models.}
    \label{fig:vis_ex_ddrh}
\end{figure*}

Analyzing the IoU scores for each category, the $\mathrm{IoU_{water}}$ for RGB images in the single model approaches saturation, whereas the $\mathrm{IoU_{water}}$ for SAR images in the single model is similar to that for RGB images. With the inclusion of the SAR modality, both CMX and ASANet exhibit improved accuracy for the water category. Notably, ASANet shows a larger increase, indicating a superior integration of SAR features. Compared to a single model using only RGB images, ASANet exhibits substantial improvements of 6.45\%, 3.33\%, and 4.61\% for the road, forest, and farmland categories, respectively, after integrating SAR features. This highlights the superior performance of ASANet in cloudy and foggy environments.

Visualization results of different models on the simulated DDHR-SK dataset with clouds are presented in Fig. \ref{fig:vis_ex_ddrh}. In this dataset, cloud occlusion reduces the texture features in RGB images, while the boundary features of roads and water remain distinct in SAR images. In visualization results, ASANet outperforms other models, producing overall superior results with smoother boundaries and fewer fine details. In Fig. \ref{fig:vis_ex_ddrh}, group a, ASANet demonstrates excellent performance by effectively integrating water features from SAR images, achieving superior detection results compared to other models. ASANet achieves higher accuracy with minimal instances of missing and false detections, distinguishing itself from CMFNet, AFNet, and CMX models. For road categories in Fig. \ref{fig:vis_ex_ddrh}, groups b, c, and d, the detection and annotation results of ASANet show relatively good overlap, further illustrating the effectiveness of ASANet in cloud-occluded scenarios.

\subsubsection{Results on WHU-OPT-SAR}\label{sec:Results on WHU-OPT-SAR}

\begin{figure*}[!h]
    \centering
    \includegraphics[width=\linewidth]{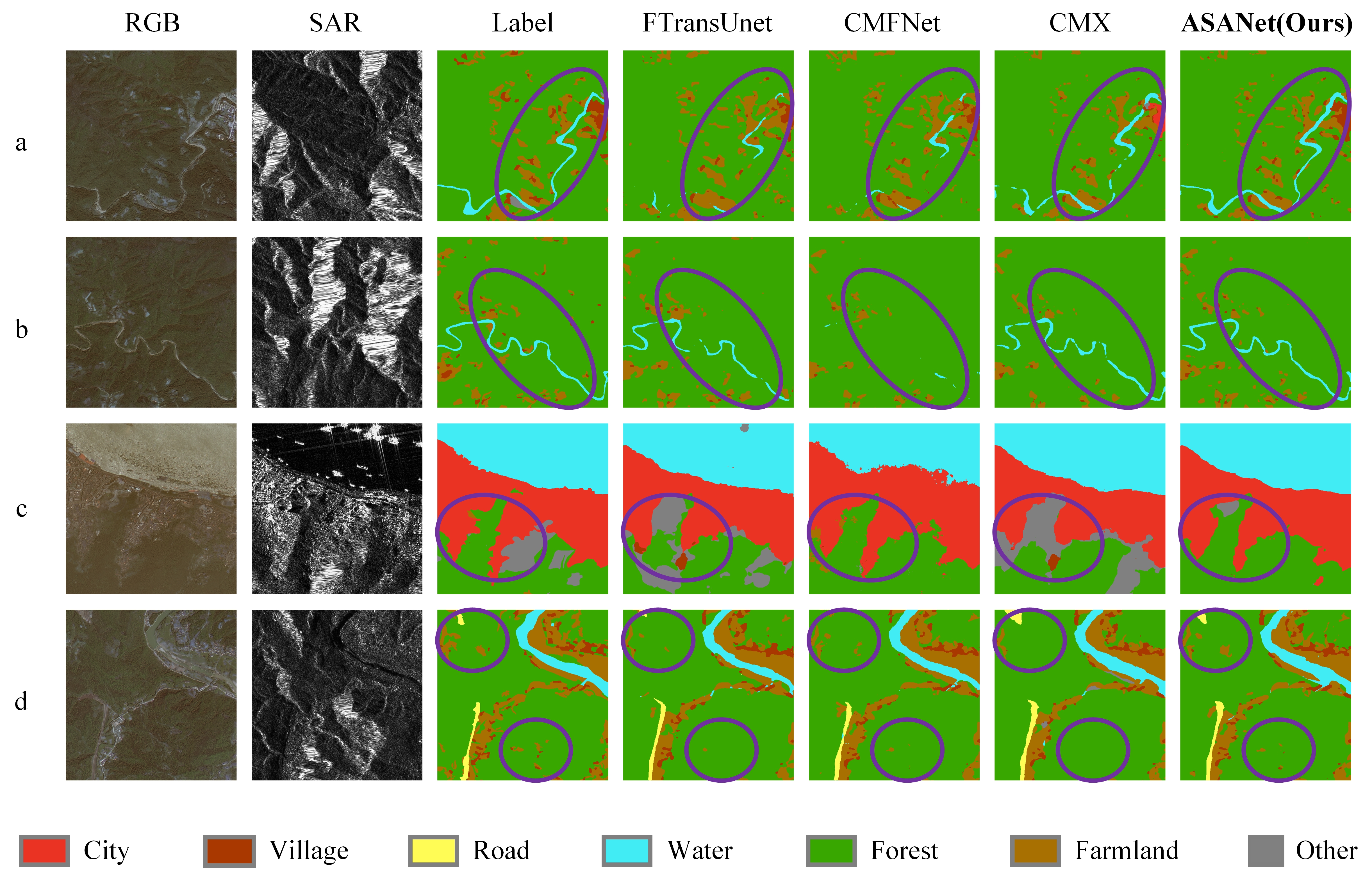}
    \caption{Visualization results of different methods on WHU-OPT-SAR dataset: four representative multimodal models.}
    \label{fig:vis_ex_whu}
\end{figure*}

The results of different models on the WHU-OPT-SAR dataset are shown in Table \ref{tab:ex_whu}. ASANet achieves the best overall results, with mIoU scores that are 15.56\%, 2.94\%, 1.64\%, 2.39\%, 2.54\%, and 0.43\% higher than those of FuseNet,  SA-Gate, FTransUNet, CMFNet, AFNet, and CMX, respectively. These results underscore the superiority of ASANet across various metrics on the WHU-OPT-SAR dataset. The visualization results are shown in the Fig. \ref{fig:vis_ex_whu}. By comparing Fig. \ref{fig:vis_ex_whu} group a and b, it can be observed that ASANet better combines the advantages of SAR images and demonstrates better connectivity in the water category. Fig. \ref{fig:vis_ex_whu} group c shows that CMFNet and our visualization method performed the best overall. However, the detection results of CMFNet contain many misdetected small patches. In Fig. \ref{fig:vis_ex_whu} group d, the ASANet model proposed by us accurately categorized the cultivated land more than other models.

\begin{table}[!h]
    \centering
    \caption{Metrics results (\%) for different methods on WHU-OPT-SAR dataset. The best results are presented in bold, while the second-best results are underlined.}
    \resizebox{1.0\linewidth}{!}{
    \begin{tabular}{ccccccccccccc}\hline
        \multirow{2}{*}{\textbf{Model}} & \multicolumn{2}{c}{\textbf{Modality}} & \multirow{2}{*}{\textbf{Kappa}} & \multirow{2}{*}{\textbf{OA}} & \multirow{2}{*}{\textbf{mIoU}} & \multicolumn{7}{c}{\textbf{IoU}}  \\ 
        & \textbf{RGB} & \textbf{SAR} & & & & \textbf{City} & \textbf{Village} & \textbf{Water} & \textbf{Road} & \textbf{Forest} & \textbf{Farmland} & \textbf{Other} \\ \hline
        UPerNet & $\checkmark$ &   & 76.13 & \uline{84.36} & \uline{55.84} & \textbf{56.80} & \textbf{49.32} & 64.56 & \uline{41.88} & 83.70 & \uline{69.95} & \uline{24.70} \\
        UPerNet &  & $\checkmark$  & 69.96 & 80.59 & 48.42 & 55.31 & 35.74 & 56.95 & 28.85 & 80.76 & 62.76 & 18.58 \\
        FuseNet & $\checkmark$ & $\checkmark$ & 65.67 & 77.67 & 38.01 & 27.94 & 36.19 & 49.35 & 13.84 & 79.55 & 59.19 & 00.00 \\ 
        SA-Gate & $\checkmark$ & $\checkmark$ & 74.13 & 83.13 & 53.17 & 53.28 & 47.39 & 62.14 & 39.19 & 82.20 & 68.05 & 19.93 \\ 
        AFNet & $\checkmark$ & $\checkmark$ & 74.96 & 83.69 & 53.57 & 54.01 & 46.93 & 62.84 & 39.41 & 83.03 & 69.06 & 19.74 \\ 
        CMFNet & $\checkmark$ & $\checkmark$ & 75.14 & 83.76 & 53.72 & 55.36 & 48.45 & 62.95 & 37.91 & 83.13 & 68.93 & 19.25 \\
        CMX & $\checkmark$ & $\checkmark$ & \uline{76.15} & \uline{84.36} & 55.68 & 56.20 & 49.02 & \uline{65.03} & 41.03 & \uline{83.72} & \uline{69.95} & \textbf{24.79} \\ 
        FTransUNet & $\checkmark$ & $\checkmark$ & 75.47 & 83.92 & 54.47 & 55.85 & 48.48 & 63.86 & 39.52 & 83.25 & 69.44 & 20.87 \\
        \textbf{ASANet(ours)} & $\checkmark$ & $\checkmark$ & \textbf{76.43} & \textbf{84.56} & \textbf{56.11} & \uline{56.79} & \uline{49.17} & \textbf{65.28} & \textbf{42.72} & \textbf{83.89} & \textbf{70.33} & 24.59 \\ \hline
    \end{tabular}
    }
\label{tab:ex_whu}
\end{table}

\section{Discussion}\label{sec:Discussion}

As described in Sec. \ref{sec:Experimental and datasets}, ASANet achieved the best results on the PIE-RGB-SAR, DDHR-SK, and WHU-OPT-SAR datasets. In this section, the paper discusses the influence of each module and the fusion of different stages on the model’s performance. Additionally, we simulate cloud and fog scenarios on RGB images and conduct comparative experiments. All settings in each experiment are consistent with those of the multimodal fusion model in the comparative experiment.

\subsection{Ablation experiments between modules}\label{sec:Ablation experiments between modules}

In this section, we conduct ablation experiments based on the PIE-RGB-SAR dataset. The results of these ablation experiments between modules are presented in Table \ref{tab:ab_pie}. Specifically, the third row of results shows the twin network framework with the SAR mode, which uses the PWA (as described in Eq. \ref{equ-6}) to fuse feature maps of the two modes before they are input into the decoder. To evaluate the effectiveness of the CFM, we replace PWA with the CFM. In assessing the effectiveness of the SFM, we integrate the SFM for feature interaction within the network framework, as shown in the third and fourth rows.

\begin{table*}[!h]
    \centering
    \caption{Metric results (\%) of ablation experiments on PIE-RGB-SAR dataset.}
    \resizebox{1.0\linewidth}{!}{
    \begin{tabular}{ccccccccccccc}\hline
        \multirow{2}{*}{\textbf{Modality}} & \multirow{2}{*}{\textbf{SFM}} & \multirow{2}{*}{\textbf{CFM}} & \multirow{2}{*}{\textbf{PWA}} & \multirow{2}{*}{\textbf{Kappa}} &\multirow{2}{*}{\textbf{OA}} & \multirow{2}{*}{\textbf{mIoU}} & \multicolumn{6}{c}{\textbf{IoU}}  \\
         & & & & &&  & \textbf{City} & \textbf{Road} & \textbf{Water} & \textbf{Forest} & \textbf{Farmland} & \textbf{Other} \\\hline
         RGB & & &  & 84.13 & 88.84 & 76.90 & 81.77 & 60.53 & 75.62 & 82.15 & 64.40 & 97.27 \\
         SAR & & &  & 74.16&83.82 & 66.73 & 75.57 & 33.69 & 70.61 & 75.76 & 49.04 & 95.73 \\ \hline
         RGB+SAR  & & & $\checkmark$ & 84.88(+0.75) & 89.36(+0.52) & 77.81(+0.91) & 82.55 & 61.50 & 76.94 & 82.90 & 65.68& 97.27 \\
         RGB+SAR  & &$\checkmark$ &  & 85.03(+0.90) & 89.47(+0.63) & 78.05(+1.15) & 82.67 & 61.63 &77.76 & 82.98 & 66.04 & 97.20 \\ \hline
         RGB+SAR & $\checkmark$ & & $\checkmark$  & 85.03(+0.15) & 89.46(+0.10) & 78.01(+0.20) & 82.58 & \textbf{61.80} & 76.78 & 83.06 & 66.52 & 97.30 \\
         \textbf{RGB+SAR} & $\checkmark$ & $\checkmark$  &  & \textbf{85.27}(+0.24) &\textbf{89.64}(+0.17) & \textbf{78.31}(+0.26) & \textbf{82.85} & \textbf{61.80} & \textbf{77.83} & \textbf{83.27} & \textbf{66.75} & \textbf{97.32} \\\hline
    \end{tabular}
    }
    \label{tab:ab_pie}
\end{table*}

\subsubsection{Analysis of results for different modes }\label{sec:Analysis of multimodal versus single-modal ablation results}

Comparing the results of the RGB-Single model (first row) with those of the SAR-Single model (second row), it is evident that the overall detection accuracy of RGB is higher than that of SAR. Notably, the $\mathrm{IoU_{road}}$ and $\mathrm{IoU_{farmland}}$ show significant differences, standing at 26.84\% and 15.36\%, respectively. The $\mathrm{IoU_{city}}$, $\mathrm{IoU_{water}}$, and $\mathrm{IoU_{forest}}$ exhibit smaller differences, at 6.2\%, 5.01\%, and 6.39\%, respectively. In the multimodal model RGB-SAR-PWA (third row), which incorporates the SAR modality, and in our proposed ASANet model, the IoU indicators for each category demonstrate improvements. These results further confirm that the texture and color features in RGB images contribute to higher accuracy, while the inclusion of SAR provides crucial invisible feature information missing in RGB. Moreover, compared to the RGB-SAR-PWA model (third row), our ASANet model (sixth row) improves mIoU by 0.5\%, Kappa by 0.39\%, and OA by 0.28\%. With the combined influence of both modules, our ASANet achieves the best IoU indicators for all categories, demonstrating superior semantic segmentation performance.

\begin{figure*}
    \centering
    \includegraphics[width=\linewidth]{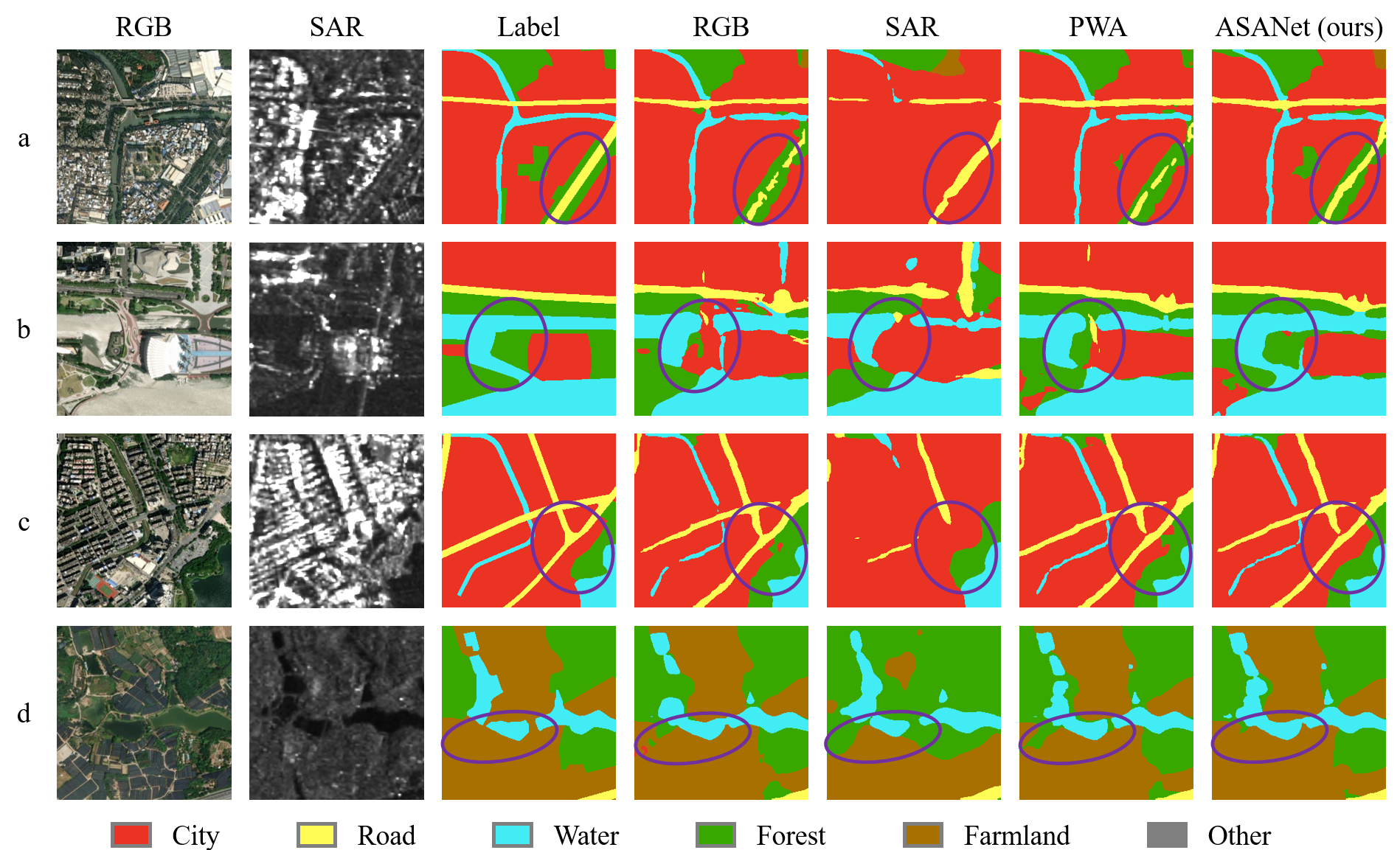}
    \caption{Visualization results of ablation experiments.}
    \label{fig:ab_pie}
\end{figure*}

In Fig. \ref{fig:ab_pie}, we visualized the results of ASANet compared to a single-modal model on the PIE-RGB-SAR dataset. The sequence, from left to right, includes the RGB image, SAR image, ground truth image, RGB-Single result, SAR-Single result, RGB-SAR-PWA result, and RGB-SAR-SFM-CFM (ASANet) result. This visualization illustrates the effectiveness of our model in utilizing both RGB and SAR modalities to enhance semantic segmentation outcomes. The analysis of the visualization results reveals that the SAR-Single model effectively captures edge features of roads and large water bodies but struggles to identify slender roads and water bodies and has difficulty distinguishing between farmland and forest. Conversely, the RGB-Single model exhibits clear texture features, resulting in accurate overall category recognition. However, they are susceptible to interference from similar texture features at the edges, resulting in the appearance of fine image spots. Specifically, observations from Fig. \ref{fig:ab_pie}, groups a (road), b (water), and c (forest) indicate that the RGB-SAR-PWA model tends to favor learning the features of RGB. In complex scenes, such as those in Fig. \ref{fig:ab_pie}, groups b and d, where texture and color features in RGB struggle to effectively classify categories, the RGB-SAR-PWA model is prone to introducing noise. For example, as evidenced by the miss segmentation of farmland as forest in Fig. \ref{fig:ab_pie} d. In contrast, our model exhibits an overall segmentation effect that mitigates the shortcomings of the two modalities. It successfully fuses complementary features from both modalities, resulting in a superior visualization effect overall. For each category, our model produces more connected segmentation results for roads and rivers and achieves greater accuracy in extracting cities, farmland, and forests.

\subsubsection{Analysis of results for different modules}\label{sec:Analysis of ablation results for different modules}

From Table \ref{tab:ab_pie}, we observe the influence of different modules on the accuracy of each category using RGB-SAR-PWA as the baseline.

\paragraph{SFM} When comparing the RGB-SAR-SFM-PWA model (fifth row) with the RGB-SAR-PWA model (third row), the most significant improvement in $\mathrm{IoU_{road}}$ and $\mathrm{IoU_{farmland}}$ is observed. In comparing the RGB-SAR-CFM model (fourth row) with the RGB-SAR-SFM-CFM model (sixth row), the enhancement in the farmland class becomes more pronounced when incorporating the SFM. Notably, these two categories are precisely those that exhibit poor performance in SAR modes. This indicates that the SFM enables the network model to flexibly adjust to feature differences between different modalities, thereby enhancing the segmentation performance for various categories.

\paragraph{CFM} The RGB-SAR-CFM model (fourth row), which integrates CFM, demonstrates improved accuracy for all foreground categories compared to the RGB-SAR-PWA model (third row). The RGB-SAR-SFM-CFM (sixth row), on the other hand, also proves the validity of CFM compared to the RGB-SAR-SFM-PWA (fifth row) model. However, when compared to the RGB-SAR-PWA (third row) model, the RGB-SAR-SFM-PWA model (fifth row) shows a slight decrease in $\mathrm{IoU_{water}}$ of 0.16\%, with the improvements in other categories being less marked. This suggests that simple feature addition does not fully harness the complementary potential of the two modalities and introduces additional noise. In contrast, the CFM we propose is specifically designed to select and fuse features from both modalities effectively.

\subsection{Ablation experiments between different stages}\label{sec:Ablation experiments between different stages}

The backbone network consists of four stages that output features at different scales: stage-1, stage-2, stage-3, and stage-4. In this section, we choose a baseline model that utilizes PWA module for fusion in all stages. Furthermore, SFM and CFM modules are introduced at various stages for ablation experiments. The fusion results for each stage are presented in Table \ref{tab:ab_stage_pie}. Our experimental findings indicate a decrease in accuracy when the SFM and CFM modules are introduced at stage-1 and stage-4. This reduction in accuracy can be attributed to the insufficient semantic information in the feature maps of the lowest stage and the limited local information in the feature maps of the highest stage. The complex feature interactions and fusions that are unique to stage-1 or stage-4 tend to amplify noise from the two modalities, thereby resulting in adverse effects.

\begin{table*}[!h]
    \centering
    \caption{Metric results (\%) of ablation experiments performed between different stages on the PIE-RGB-SAR dataset.}
    \resizebox{1.0\linewidth}{!}{
    \begin{tabular}{ccccccccccccc}\hline
        \multicolumn{4}{c}{\textbf{Stage}} &\multirow{2}{*}{\textbf{Kappa}} & \multirow{2}{*}{\textbf{OA}} & \multirow{2}{*}{\textbf{mIoU}} & \multicolumn{6}{c}{\textbf{IoU}}  \\ 
         \textbf{1} & \textbf{2} & \textbf{3} & \textbf{4} & & & & \textbf{City} & \textbf{Road} & \textbf{Water} & \textbf{Forest} & \textbf{Farmland} & \textbf{Other}\\ \hline
         &  &  &  & 84.88 & 89.36 & 77.81 & 82.55 & 61.5 & 76.94 & 82.9 & 65.68 & 97.27 \\  \hline
        $\checkmark$&  &  &  & 84.85(-0.03) & 89.35(-0.01) & 77.80(-0.01) & 82.55 & 61.62 & 76.92 & 82.86 & 65.51 & 97.31 \\ 
        & $\checkmark$ &  &  &\textbf{85.09}(+0.21)& \textbf{89.51}(+0.15) & \textbf{78.10}(+0.29) & 82.70 & 61.77 & \textbf{77.50} & \textbf{83.07} & \textbf{66.31} & 97.27 \\ 
        &  &  $\checkmark$ &  & 84.97(+0.09) & 89.43(+0.07) & 77.92(+0.11) & \textbf{82.77} & \textbf{61.93} & 76.94 & 82.98 & 65.52 & \textbf{97.37} \\ 
        &  &  & $\checkmark$  & 84.52(-0.36) & 89.11(-0.25) & 77.38(-0.43) & 82.17 & 61.15 & 76.22 & 82.53 & 64.92 & 97.31 \\  \hline
        $\checkmark$ & $\checkmark$ & $\checkmark$ &  $\checkmark$  &\textbf{85.27}(+0.39)& \textbf{89.64}(+0.28) & \textbf{78.31}(+0.5) & \textbf{82.85} & 61.80 & \textbf{77.83} & \textbf{83.27} & \textbf{66.75} & 97.32 \\ \hline
    \end{tabular}
    }
\label{tab:ab_stage_pie}
\end{table*}

The outcomes of incorporating SFM and CFM across different stages indicate that water bodies, forests, and croplands exhibit superior performance at stage-2, with roads and cities performing optimally at stage-3. The local details and texture features in stage-2 benefit significantly from integration. For categories that require expansive receptive fields and rich semantic information, stage-3 offers the most effective fusion. However, at stage-4, the drop in accuracy is more significant, due to the reduced availability of location-specific information.

The network that integrated SFM and CFM modules across all stages achieved the highest overall accuracy. However, the $\mathrm{IoU_{road}}$ and $\mathrm{IoU_{other}}$ decreased compared to the network where only the stage-3 was enhanced. This suggests that features fused across different stages may negatively impact certain categories. For detection tasks requiring high accuracy for specific categories, combining fusion strategies across different stages can yield superior results.

\subsection{Comparative experiments under cloudy conditions}\label{sec:Comparative experiments under fog conditions}

\begin{figure}[!h]
    \centering
    \includegraphics[width=\linewidth]{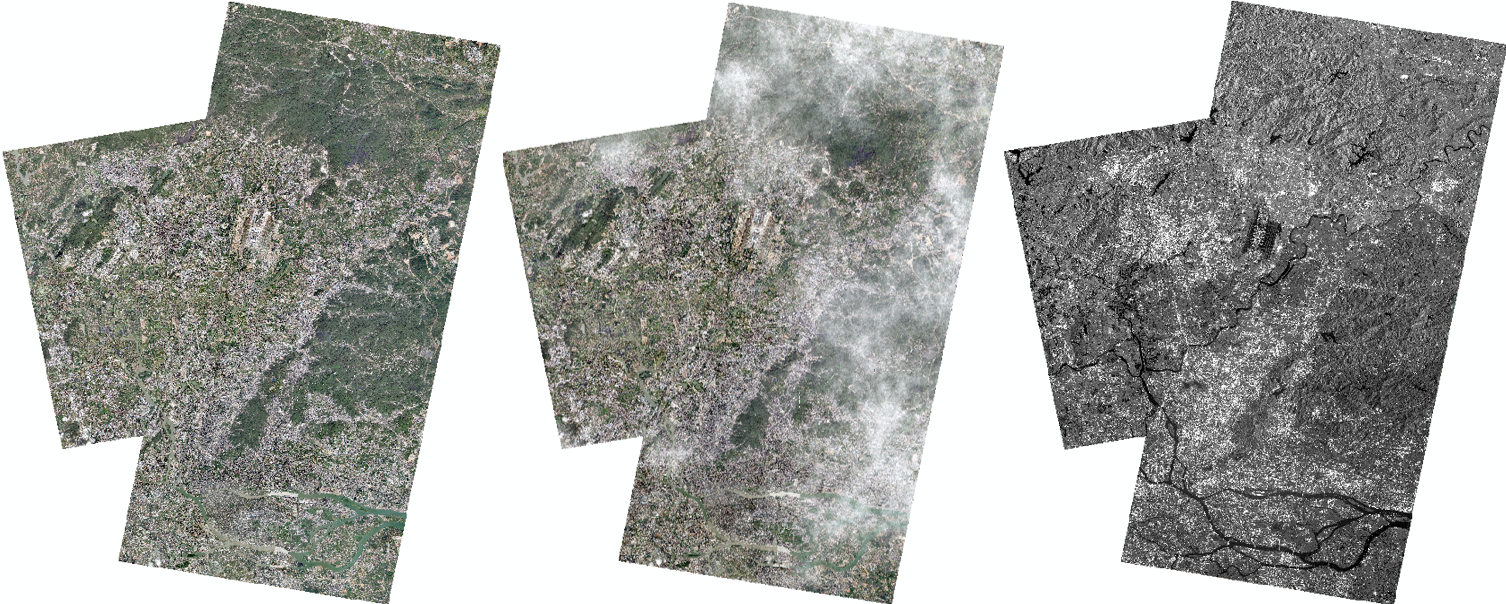}
    \caption{Overall image of the simulated cloudy scene: RGB on the left, Simulated RGB in the middle, and SAR on the right}
    \label{fig:clouds}
\end{figure}

To assess the model’s effectiveness and superiority in capturing large-scale cloudy scenes, we generated RGB images under simulated cloudy conditions using GIMP. (The foggify filter was used to perform the simulation, with fog color set to white, turbulence set to 1.5, and opacity set to 100.) The simulated image is depicted in Fig. \ref{fig:clouds}. Adhering to the methodology employed in PIE-RGB-SAR, we cropped the images and segmented the dataset. The experimental outcomes for diverse models are summarized in Table \ref{tab:foggy}.

\begin{table}[!h]
    \centering
    \caption{Comparative experimental metrics (\%) of different models in a cloudy scene. The best results are presented in bold, while the second-best results are underlined.}
    \resizebox{1.0\linewidth}{!}{
    \begin{tabular}{cccccccccccc}\hline
        \multirow{2}{*}{\textbf{Model}} & \multicolumn{2}{c}{\textbf{Modality}} &\multirow{2}{*}{\textbf{Kappa}} & \multirow{2}{*}{\textbf{OA}} & \multirow{2}{*}{\textbf{mIoU}} & \multicolumn{6}{c}{\textbf{IoU}} \\ 
        & \textbf{RGB} & \textbf{SAR} & & & & \textbf{City} & \textbf{Road} & \textbf{Water} & \textbf{Forest} & \textbf{Farmland} & \textbf{Other}\\ \hline
        UPerNet & $\checkmark$ & & 81.95 & 87.31 & 73.18 & 79.68 & 50.49 & 68.03 & 80.38 & 63.99 & 96.51 \\ 
        UPerNet & & $\checkmark$ & 74.16 & 83.82 & 66.73 & 75.59 & 33.74 & 70.42 & 75.98 & 49.60 & 95.74 \\ 
        FuseNet & $\checkmark$ & $\checkmark$ & 66.40 & 77.14 & 54.60 & 67.96 & 19.40 & 49.24 & 67.85 & 27.04 & 96.10 \\ 
        SA-Gate & $\checkmark$ & $\checkmark$ & 79.34 & 85.54 & 69.21 & 77.67 & 39.41 & 64.02 & 78.01 & 59.73 & 96.44 \\
        AFNet & $\checkmark$ & $\checkmark$ & 81.40& 86.91 & 72.00 & 79.43 & 45.89 & 69.17 & 80.04 & 62.21 & 95.29\\ 
        CMFNet & $\checkmark$ & $\checkmark$ & 81.78 & 87.22 & 72.77 & 80.4 & 48.15 & 73.11 & 80.15 & 58.17 & 96.63 \\
        CMX & $\checkmark$ & $\checkmark$ & \uline{83.52} & \uline{88.43} & \uline{75.13} & \uline{81.51} & \uline{53.06} & 72.97 & \uline{81.89} & \uline{64.20} & \textbf{97.13} \\ 
        FTransUNet & $\checkmark$ & $\checkmark$ & 81.70 & 87.13 & 72.78 & 80.24 & 48.09 & \uline{73.61} & 80.06 & 57.75 & 96.89 \\
        \textbf{ASANet(ours)} & $\checkmark$ & $\checkmark$ & \textbf{84.15} & \textbf{88.85} & \textbf{76.20} & \textbf{81.76} & \textbf{55.34} & \textbf{74.46} & \textbf{82.48} & \textbf{66.09} & \uline{97.07} \\  \hline
    \end{tabular}
    }   
\label{tab:foggy}
\end{table}

When compared to the original RGB image, the kappa, OA, and mIoU of the simulated S-RGB single-mode image decreased by 2.18\%, 1.53\%, and 3.72\%, respectively. This suggests that foggy conditions can adversely affect detection outcomes. Nevertheless, in the simulated environment, our ASANet achieved the best performance, with the mIoU index 1.07\% to 21.06\% higher than the corresponding indices of other models. The advantage of ASANet is more obvious in the road and water categories.

\section{Conclusion}\label{sec:Conclusion}

In this study, we introduce ASANet, which asymmetrically fuses RGB and SAR images at the feature level for LCC. We design an SFM module to asymmetric enhance the interaction between features of the two modalities, adaptively amplifying complementary features in distinct branches. Additionally, we introduce a CFM to perform feature fusion by selectively integrating features from both modalities in both channel and spatial dimensions. Furthermore, we introduce a multimodal LCC dataset that is designed for practical application scenarios. We conduct extensive experiments on this dataset as well as two other publicly available datasets, comparing our ASANet with existing multimodal fusion methods. The experimental results show that our ASANet achieves excellent performance in LCC under both real-world and simulated cloud and fog conditions. The effectiveness of our model is also evident in feature map visualizations, which highlight the model’s ability to adaptively utilize information from both modalities, mitigate unimodal limitations, and achieve complementary information fusion. In our future work, we will focus on investigating the application of an asymmetric modal fusion architecture. This approach aims to align the modal information with the corresponding parameters of each branch, thereby optimizing the utilization of modal characteristics.

\section*{Declaration of competing interest}\label{Declaration of competing interest}

The authors declare that they have no known competing financial interests or personal relationships that could have appeared to influence the work reported in this paper.



\bibliographystyle{elsarticle-harv} 
\bibliography{cas-refs}





\end{document}